\documentclass[sigconf]{acmart}

\usepackage{enumitem} 
\usepackage{tabularx} 
\usepackage{siunitx} 
\usepackage{makecell}

\AtBeginDocument{%
  \providecommand\BibTeX{{%
    \normalfont B\kern-0.5em{\scshape i\kern-0.25em b}\kern-0.8em\TeX}}}

\copyrightyear{2026}
\acmYear{2026}
\setcopyright{cc}
\setcctype{by-nc-nd}
\acmConference[UIST '26]{The 39th Annual ACM Symposium on User Interface Software and Technology}{November 02--05, 2026}{Detroit, MI, USA}
\acmBooktitle{The 39th Annual ACM Symposium on User Interface Software and Technology (UIST '26), November 02--05, 2026, Detroit, MI, USA}
\acmDOI{10.1145/3830398.3830562}
\acmISBN{979-8-4007-2856-3/2026/11}

\begin{document}

\title[InvisIto]{InvisIto: Weaving Unobtrusive Infrared Markers\\for Ubiquitous Textile Interaction}

\author{Hsuanling Lee}
\email{hsuanling.lee@utdallas.edu}
\authornote{Both authors contributed equally to this research.}
\authornote{Also with Keio University.}  
\orcid{0009-0000-4832-0458}
\affiliation{%
  \institution{University of Texas at Dallas}
  \city{Richardson}
  \state{TX}
  \country{USA}
}

\author{Hal Sugiyama}
\authornotemark[1] 
\email{hal@pplab.jp}
\orcid{0009-0009-5480-5991}
\affiliation{%
  \institution{Keio University}
  \city{Kanagawa}
  \country{Japan}
}

\author{Tian Min}
\email{welkinmin@keio.jp}
\orcid{0000-0003-0912-4598}
\affiliation{%
 \institution{Keio University}
  \city{Yokohama}
  \country{Japan}
}

\author{Hanako Fujino}
\email{hhf@pplab.jp}
\orcid{0009-0007-9388-6224}
\affiliation{%
 \institution{Keio University}
  \city{Kanagawa}
  \country{Japan}
}

\author{Mayuka Kuwana}
\email{mkuwana@g.ecc.u-tokyo.ac.jp}
\orcid{0009-0008-9832-5490}
\affiliation{%
 \institution{The University of Tokyo}
  \city{Tokyo}
  \country{Japan}
}

\author{Yuta Sugiura}
\email{sugiura@keio.jp}
\orcid{0000-0003-3735-4809}
\affiliation{%
 \institution{Keio University}
  \city{Yokohama}
  \country{Japan}
}

\author{Mustafa Doga Dogan}
\email{doga@adobe.com}
\orcid{0000-0003-3983-1955}
\affiliation{%
  \institution{Adobe Research}
  \city{Basel}
  \country{Switzerland}
}

\author{Liang He}
\authornotemark[2] 
\email{liang.he@utdallas.edu}
\orcid{0000-0003-4826-629X}
\affiliation{%
  \institution{University of Texas at Dallas}
  \city{Richardson}
  \state{TX}
  \country{USA}
}

\author{Koya Narumi}
\email{narumi@keio.jp}
\orcid{0000-0001-5830-3942}
\affiliation{%
  \institution{Keio University}
  \city{Kanagawa}
  \country{Japan}
}

\renewcommand{\shortauthors}{Lee and Sugiyama, et al.}

\begin{abstract}
Textiles are increasingly explored as media for interacting with digital information. However, many of the existing approaches rely on visible tags, printed overlays, or electronic modules that compromise the fabric’s aesthetic and tactile qualities.
To address this, we present \textit{InvisIto}, a method for weaving visually unobtrusive yet machine-readable infrared markers directly into fabrics using near-infrared (NIR)-absorbing yarns. 
Although these yarns look similar to standard fibers in ambient light, they produce strong contrast in NIR imaging.
Our method includes: (1) a design tool that helps users easily embed infrared markers into weaving drafts, (2) five disguising strategies that further reduce marker visibility under ambient light, and (3) a camera-based detection pipeline for decoding and tracking the woven markers. 
InvisIto supports both woven QR codes for data encoding and woven ArUco markers for binary input and deformation tracking.
We demonstrate applications across hand weaving, Jacquard weaving, and industrial fabrication, showing that InvisIto supports interaction and fabrication across scales, from bespoke artifacts to mass production.
\end{abstract}





\begin{teaserfigure}
    \centering
    \includegraphics[width=\textwidth]{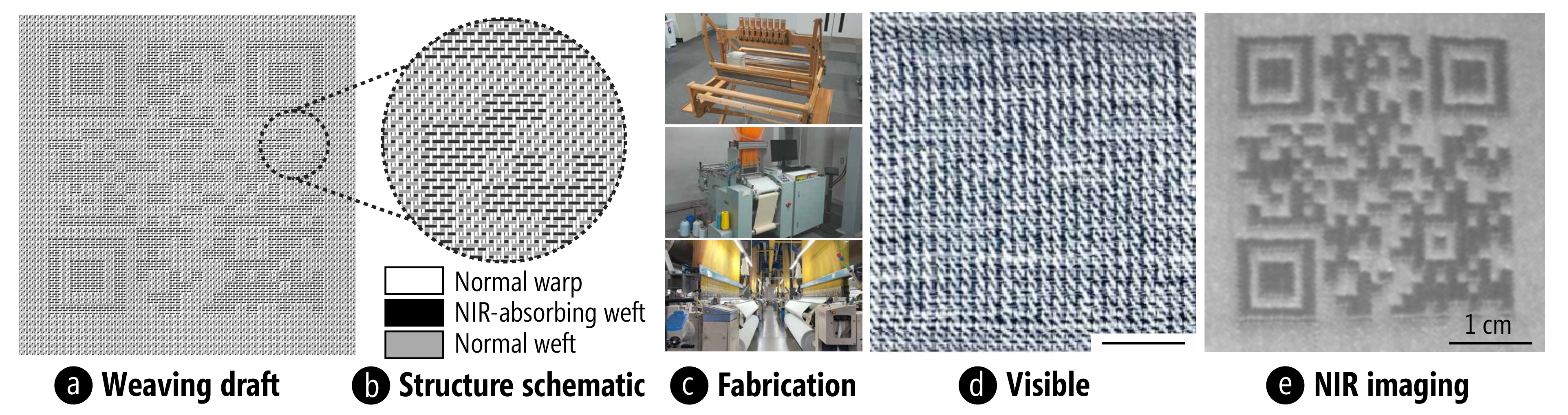}
    \caption{\textbf{InvisIto overview.} (a) A woven QR code is authored as a weaving draft. (b) Marker contrast is produced through yarn-level structure design using a normal warp, a normal weft, and an NIR-absorbing weft. (c) The markers can be fabricated across multiple weaving contexts, including hand weaving, Jacquard weaving, and industrial textile production. (d) With visible imaging, the woven marker remains visually unobtrusive within the textile surface. (e) With near-infrared (NIR) imaging, the embedded QR code pattern becomes clearly machine-readable.}
    \Description{Five-part figure showing the InvisIto workflow and result. A weaving draft containing a QR code-like pattern is shown first, followed by an enlarged schematic of the woven structure using normal warp, NIR-absorbing weft, and normal weft yarns. The figure then shows fabrication across hand weaving, Jacquard weaving, and industrial weaving, and ends with a comparison of the same textile under visible light and near-infrared imaging, where the marker is faint in visible light but clearly revealed in NIR.}
    \label{fig:teaser}
\end{teaserfigure}
\maketitle

\section{Introduction}
\label{sec:introduction}
Textiles have long served as media for encoding and communicating information, from the ancient Inca \textit{quipu} cords used for data recording to the Jacquard loom's punched-card weaving system that inspired early computing. 
This legacy continues in human-computer interaction (HCI), where researchers explore textiles as substrates for digital-physical integration, embodying Mark Weiser's vision of \textit{ubiquitous computing}, where technologies ``weave themselves into the fabric of everyday life''~\cite{weiser_computer_1999}.

Many current approaches to integrating textiles with digital information and interaction, however, rely on visible components~\cite{chiu_fabric_2025,k_p_weaving_2023,haberfellner_threads_2024,preston_enabling_2017}, embedded electronics~\cite{pouta_woven_2022,devendorf_i_2016}, or additive layers~\cite{nofal_initiating_2020}.
Although textiles are well-suited to ubiquitous computing because of their intimate integration with our everyday activities, current approaches often undermine what makes them appealing in the first place, including their aesthetics, comfort, and ability to blend into daily life. 
For example, visible markers disrupt visual design, embedded electronics can alter flexibility and washability, and surface treatments may wear out over time or change how the fabric feels.

Marker systems offer a simple and passive means of adding digital functionality to textiles. They have been used to support data encoding~\cite{nofal_initiating_2020}, identification~\cite{haberfellner_threads_2024,chiu_fabric_2025}, and tracking~\cite{liang_chic-marker_2024, Zheng2020, Xiao2025} in textile contexts. However, most existing implementations remain visually intrusive. As a promising alternative to these markers, HCI researchers have investigated infrared (IR) markers across various substrates that are readable only at wavelengths less salient to human vision~\cite{dogan_brightmarker_2023,dogan_infraredtags_2022,feick_imprinto_2025}. 
Inspired by this line of work, our goal is to achieve such IR marker functionality on fabric substrates.
Although some related works in materials science share a similar motivation and have explored infrared functionality in textile-related systems~\cite{barcode,iezzi_fiber_2024}, their primary focus is on novel material development rather than on scalable deployment or interaction design. 


We present \textit{InvisIto}\footnote{The method name \textit{InvisIto} combines \textit{Invisible} and \textit{Ito} (meaning ``yarn'' in Japanese).}, a method for weaving visually unobtrusive yet machine-readable markers directly into textiles using NIR-absorbing yarn (\autoref{fig:teaser}).
This work extends our prior work~\cite{sugiyama_weaving_2026}, which introduced the concept of weaving NIR-absorbing yarn into fabric to create hidden QR codes and explored early methods for reducing their visibility.
InvisIto further provides an end-to-end workflow for weaving markers with NIR-absorbing yarn, including (1) a design tool for accessibly designing and weaving NIR markers on textiles, (2) five disguising strategies that further reduce marker visibility under ambient light, and (3) a camera-based detection pipeline for decoding and tracking the woven NIR markers.
The resulting fabrics remain visually unobtrusive under visible light while revealing high-contrast marker patterns under NIR imaging. InvisIto supports two classes of functionality: (1) digital information encoding through woven QR codes and (2) binary input and real-time deformation tracking through woven ArUco markers.

We further demonstrate these capabilities across multiple fabrication contexts (\autoref{fig:teaser}c), from hand weaving for bespoke artifacts to industrial Jacquard production for scalable manufacturing. Our example applications include an anti-counterfeit T-shirt with embedded identity markers, a hand-woven wristband encoding personal narratives, an industrially fabricated sofa upholstery, and an AR-enhanced cushion that supports real-time deformation tracking.

To summarize, our work makes the following contributions:
\begin{enumerate}[label=(\arabic*)] 
    \item \textbf{Yarn-level unobtrusive marker integration:} a method for weaving visually unobtrusive yet machine-readable NIR markers directly into fabrics;
    \item \textbf{Marker authoring and disguising workflow:} a design tool for integrating markers into weaving drafts, together with five disguising strategies that reduce marker visibility under visible light while preserving NIR readability;
    \item \textbf{Detection pipeline:} an NIR sensing pipeline enabling recognition of static markers, binary input, and real-time tracking of textile deformation; and
    \item \textbf{Applications across fabrication scales:} applications spanning hand weaving, Jacquard weaving, and industrial textile production.
\end{enumerate}

\section{Related Work}
\label{sec:relatedwork}
Our work builds upon three strands of prior research: infrared-based marker systems, textile-based data encoding, and textile deformation sensing.


\subsection{Infrared-Based Marker Systems}
A persistent challenge in marker-based interaction with physical objects is that visible markers disrupt aesthetics and occupy visual space. 
In HCI, infrared (IR) markers have been increasingly explored to address this. These systems exploit material properties such as IR absorption~\cite{willis_hideout_2013, feick_imprinto_2025, punpongsanon_deforme_2013}, transmission~\cite{dogan_infraredtags_2022}, or fluorescence~\cite{dogan_brightmarker_2023, silapasuphakornwong_embedding_2020} to produce patterns that are unobtrusive to the human eye yet readable by machines. They have supported applications such as AR tracking~\cite{dogan_brightmarker_2023, dogan_standarone_2023}, projection mapping~\cite{tone_fibar_2020}, and personalized interaction~\cite{rosner_spyn_2010}.

Many prior implementations target rigid or semi-rigid substrates.
For example, \textit{InfraredTags}~\cite{dogan_infraredtags_2022} encodes IR-transmissive barcodes inside 3D-printed plastics to support low-cost, invisible object tagging. 
\textit{BrightMarker}~\cite{dogan_brightmarker_2023} incorporates NIR-fluorescent filaments into 3D-printed structures, enabling high-contrast marker tracking under motion and ambient light. 
Research such as \textit{Imprinto}~\cite{feick_imprinto_2025} and \textit{StandARone}~\cite{dogan_standarone_2023} explores the use of off-the-shelf IR-absorbing inks for inkjet printing, enabling the creation of invisible watermarks or data patterns on documents. 
However, direct textile integration remains rare. \citeauthor{rosner_spyn_2010}~\cite{rosner_spyn_2010} surface-painted IR-absorbing ink onto finished knit fabric for location-aware annotation, though surface treatments wear with use and sit outside the garment's production process.

Related work in materials science has also explored encoding IR markers in textiles. \citeauthor{barcode}~\cite{barcode} constructed an IR-emitting barcode from an array of conductive yarns that require an applied voltage to function, while \citeauthor{iezzi_polymeric_2023}~\cite{iezzi_polymeric_2023} developed photonic-crystal fibers that reflect specific infrared wavelengths through custom fiber synthesis. Although these studies confirm the feasibility of IR encoding in fibrous structures, they prioritize material development over practical deployment in HCI contexts and across real-world production scales.

\subsection{Textile-Based Data Encoding}
The ability to encode data into textiles transforms them from passive materials into information-carrying media. 
Prior HCI research has explored embedding digital information in textiles for supply chain traceability~\cite{haberfellner_threads_2024, iezzi_polymeric_2023, iezzi_fiber_2024}, interactive experiences~\cite{chan_data_2017, rosner_spyn_2010}, and sustainable material practices~\cite{chiu_fabric_2025}. Early approaches augmented fabrics with external or surface-level encodings.
For example, RFID tags have been affixed to garments for logistics and inventory tracking~\cite{jankowski-mihulowicz_identification_2023}, and sewn wash tags or printed QR codes have been used to provide readable product information. 
These commercially oriented tagging techniques inspired a range of HCI research that adapted similar principles for interactive and expressive applications in garments.
\citeauthor{hakkila_clothing_2017}~\cite{hakkila_clothing_2017} attached AR markers into clothing patterns to enable smartphone-mediated interactions, while \citeauthor{kuusk_crafting_nodate}~\cite{kuusk_crafting_nodate} introduced QR-coded embroidery to link craft-based textiles with digital narratives. 
These studies laid important groundwork for data encoding in textiles. However, they frequently relied on add-ons that visibly altered the textile.

More recent work embeds data through the fabric's structure itself, including magnetized weave patterns~\cite{chan_data_2017}, woven identifiers~\cite{haberfellner_threads_2024}, weave-based AR fiducials~\cite{k_p_weaving_2023}, and Jacquard-woven conductive elements~\cite{pouta_woven_2022}.
While these advances improve structural integration, they often restrict design choices, require post-processing, or use materials that reduce softness and wearability.

\subsection{Textile-Based Deformation Tracking}
Deformation tracking in textiles has evolved along two primary paths in HCI. Sensor-based methods integrate conductive materials~\cite{yu_seampose_2024, luo_knitui_2021, gioberto_detecting_2013, ray_capafoldable_2023} or functional coatings~\cite{li_e-textile_2024, peng_intellitex_2024, honnet_polysense_2020} to translate mechanical deformation into electrical signals. While these techniques are relatively accurate, they often introduce non-textile elements, such as conductive layers, threads, or coatings, onto the original fabric surface and require additional wiring, power, and data-acquisition hardware to read the sensor signals.
These additions often complicate integration and compromise the textile’s inherent softness and breathability. \citeauthor{xu_bit_2025}~\cite{xu_bit_2025} aim to solve this by using near-field coupling. However, the need for specialized electronic integration remains.

Vision-based methods instead infer deformation from optical cues such as fiducial markers~\cite{liang_chic-marker_2024, yaldiz_deepformabletag_2021, zhu_research_2023, maida_claycode_2025, steimle_flexpad_2013} or embedded patterns~\cite{narita_dynamic_2015, halimi_garment_2022}. While these approaches avoid rigid electronics, they still rely on visible surface features that can compete with the textile’s intended visual appearance.
\textit{Chic-Marker}~\cite{liang_chic-marker_2024}, for example, computationally fuses \textit{AprilTag} into a \textit{Pied-de-poule} pattern via sublimation printing, reducing the visual salience of the marker. However, it still does not resolve the fundamental reliance on visible surface markings.
These approaches still rely on visible markers or surface patterns, leaving unobtrusive textile deformation tracking an open challenge.

Taken together, InvisIto occupies a distinct position among existing approaches by integrating visually unobtrusive, passive, and machine-readable markers directly into the woven structure of textiles, enabling both data encoding and deformation tracking within standard textile fabrication workflows. 
Rather than focusing on rapid prototyping through additive or surface-applied fabrication techniques, InvisIto is designed to integrate with existing weaving workflows, supporting scalable textile manufacturing~\cite{poupyrev_project_2016,zebrasense}. 
Unlike surface-applied, embroidered, electronic, or conductive alternatives, InvisIto combines washability, scalable textile manufacturing, and deformation tracking without requiring embedded electronics or obvious surface graphics.
\section{InvisIto}
\label{sec:iryarn}


\subsection{Overview}
InvisIto embeds visually unobtrusive but machine-readable markers into woven textiles using NIR-absorbing yarns. The workflow consists of three stages: marker design and embedding through our design tool, textile fabrication through weaving, and marker decoding or tracking under NIR imaging. 
\begin{figure}[H]
    \centering
    \includegraphics[width=\linewidth]{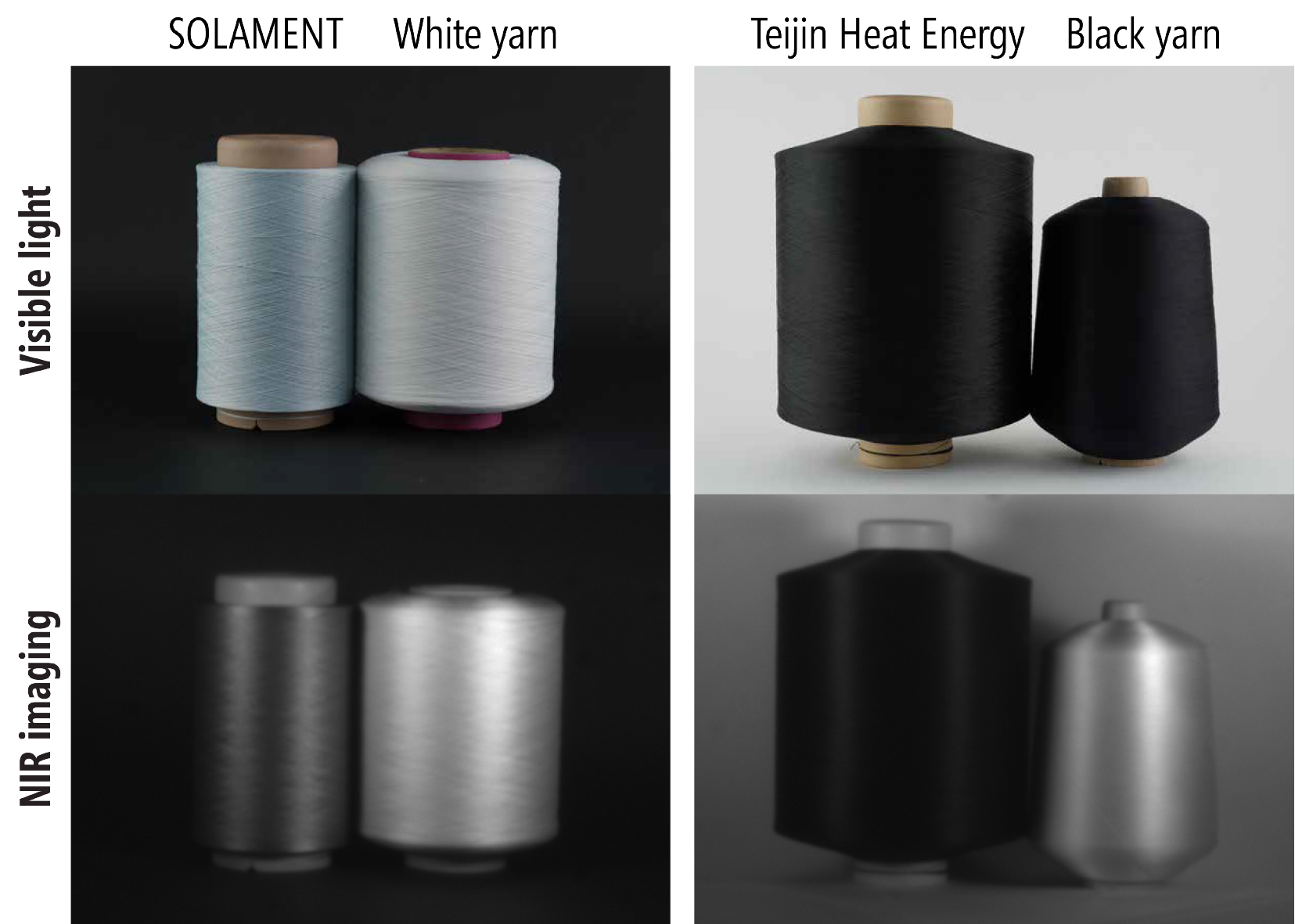}
    \caption{Visible-light and NIR images of SOLAMENT (8\%) and Teijin Heat Energy yarns compared with normal white and black yarns, respectively. }
    \Description{A two-row comparison of four yarn cones. The top row shows SOLAMENT beside normal white yarn and Teijin Heat Energy beside normal black yarn under visible light. SOLAMENT appears similar to the white yarn, while Teijin Heat Energy appears similar to the black yarn. In the bottom row, under NIR imaging, SOLAMENT and Teijin Heat Energy appear dark, whereas the normal white and black yarns appear bright, producing clear contrast between each pair.}
    \label{fig:yarn}
\end{figure}
\subsection{NIR-Absorbing Yarns}
Near-infrared (NIR)-absorbing yarns serve as the material basis of InvisIto. NIR yarns are engineered to selectively attenuate light in the near-infrared (NIR) range (approximately \qtyrange{700}{1400}{\nano\meter}) while remaining visually neutral under visible light. As shown in \autoref{fig:yarn}, these yarns closely match standard yarns under ambient light yet appear darker under NIR imaging, which enables us to embed machine-readable contrast directly into the textile structure rather than adding visible tags, printed overlays, or attached modules.
In our implementations, we used two types of commercially available IR-absorbing polyester yarns: SOLAMENT\footnote{\url{https://crossmining.smm-g.com/solament/}} MKT2205A (\qty{150}{D}, Sumitomo Metal Mining\footnote{\url{https://www.smm.co.jp/en/}}), which is available in variants with 2\%, 4\%, 6\%, and 8\% NIR-absorbing additive content, and Teijin Heat Energy TWYCKU160 BCE167T48 (\qty{150}{D}).

\begin{figure}[h]
    \centering
    \includegraphics[width=\linewidth]{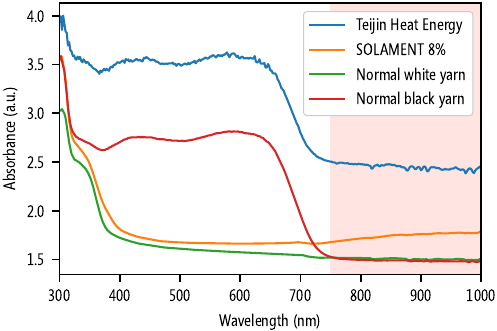} 
    \caption{Absorbance spectra of SOLAMENT, Teijin Heat Energy, regular white yarn, and regular black yarn measured from \qtyrange[mode=text,reset-text-family=false,reset-text-series=false,reset-text-shape=false,range-phrase={ to }]{300}{1000}{\nano\meter}. The shaded region indicates the near-infrared range from \qtyrange[mode=text,reset-text-family=false,reset-text-series=false,reset-text-shape=false,range-phrase={ to }]{750}{1000}{\nano\meter}.}
    \Description{Line plot comparing yarn absorbance across the near-infrared range. The x-axis shows wavelength from 300 to 1000 nanometers, and the y-axis shows absorbance in arbitrary units. 
    }
    \label{fig:spectrum}
\end{figure}

To characterize the NIR absorbance of the yarn materials used in InvisIto, \autoref{fig:spectrum} compares the absorbance spectra of SOLAMENT (8\%), Teijin Heat Energy, and normal yarns of white and black color measured with a spectrophotometer (HORIBA Duetta\footnote{\url{https://www.horiba.com/int/scientific/products/detail/action/show/Product/duetta-1621/}}). 
Although the absorbance properties are similar in the visible region between the normal yarn and the corresponding NIR-absorbing yarn, both NIR-absorbing yarns exhibit consistently higher absorbance than normal yarns across the NIR region from \qtyrange[range-phrase={ to }]{750}{1000}{\nano\meter}, confirming their stronger NIR attenuation.

\subsection{Embedding Markers}
Weaving interlaces two sets of yarns: warp yarns run lengthwise and stay fixed on the loom, while weft yarns are widthwise inserted across them to form the textile structure. InvisIto encodes a marker with two wefts that look similar in ambient light but differ under NIR imaging: an NIR-absorbing yarn and a regular (i.e., NIR-transparent) yarn. The warp is entirely standard yarn; the weft alternates between the two types. Under NIR imaging, the NIR-absorbing weft reads as dark and the standard weft as bright, giving the ``black'' and ``white'' of a binary marker.
\begin{figure}[th]
  \centering
  \includegraphics[width=\linewidth]{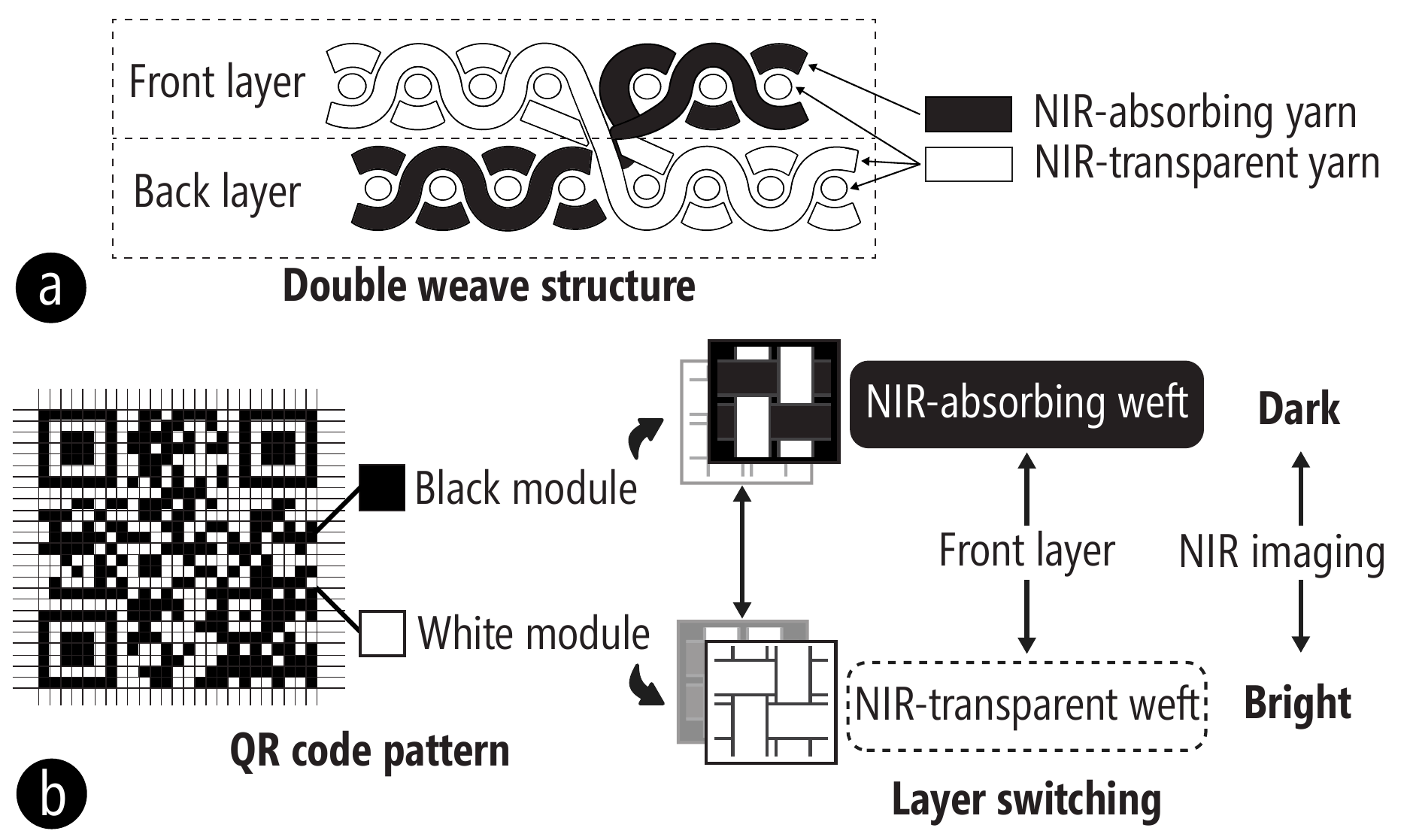}
  \caption{Double-weave encoding of QR modules. Layer switching brings either the NIR-absorbing or NIR-transparent weft to the front surface, producing dark and bright modules under NIR imaging for woven QR code generation.}
  \Description{Diagram of the double-weave encoding mechanism used in InvisIto. The upper portion shows a cross-sectional schematic and 3D renderings of a double-weave textile with front and back layers. The lower portion shows a QR code pattern composed of black and white modules, alongside two weave states produced by layer switching. In one state, the NIR-absorbing weft is brought to the front layer and appears dark under NIR imaging. In the other, the NIR-transparent weft is brought to the front layer, where it appears bright. Arrows indicate the correspondence between QR code modules, layer assignment, and NIR appearance.}
  \label{fig:double_weave}
\end{figure}

\begin{figure*}[tbp]
  \centering
  \includegraphics[width=\linewidth]{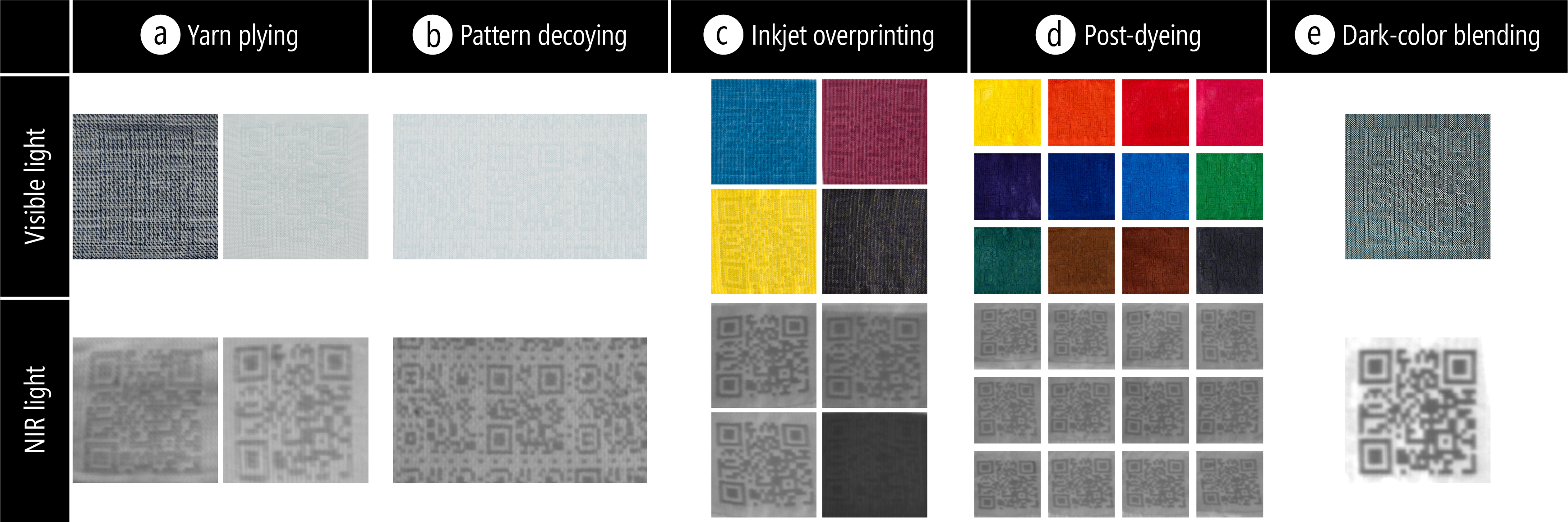}
    \caption{Comparison of five disguising strategies: (a) yarn plying with two colors of coiling yarns (navy and white), (b) pattern decoying, (c) inkjet overprinting with CMYK inks, (d) post-dyeing with 12 different colors, (e) dark-color blending. The photos were taken under visible light (top) and under NIR light (bottom).}
    \Description{Five-column comparison figure. Each column shows one disguising strategy, with a visible-light photo on top and a near-infrared photo below. From left to right: (a) yarn plying using navy and white coiling yarns, (b) pattern decoying, (c) inkjet overprinting with CMYK inks, (d) post-dyeing in 12 different colors, and (e) dark-color blending.}
  \label{fig:disguise}
\end{figure*}

To control which weft yarn is visible at each location on the fabric surface, we use double weave. Double weave is a weaving technique to produce two fabric layers, a Front layer and a Back layer,  simultaneously on the same loom (\autoref{fig:double_weave}a). 
The marker pattern is created by switching the two weft yarns between these layers.
When the NIR-absorbing weft is brought to the Front layer and the standard weft is routed to the Back layer, the surface reads as dark under NIR imaging. Swapping the two layers reverses this, producing a bright region (\autoref{fig:double_weave}b). By applying this layer-switching operation module-by-module, we encode the complete binary marker. 
Our design tool generates the double weave structure for marker design as weaving drafts (\autoref{fig:teaser}a).  For each weft insertion, or pick, the draft specifies which warp yarns the loom raises, which determines how the NIR-absorbing and standard wefts are assigned to the Front and Back layers, thereby encoding the marker through double weave.

\subsection{Disguising Markers}
\begin{figure}[htbp]
  \centering
  \includegraphics[width=\linewidth]{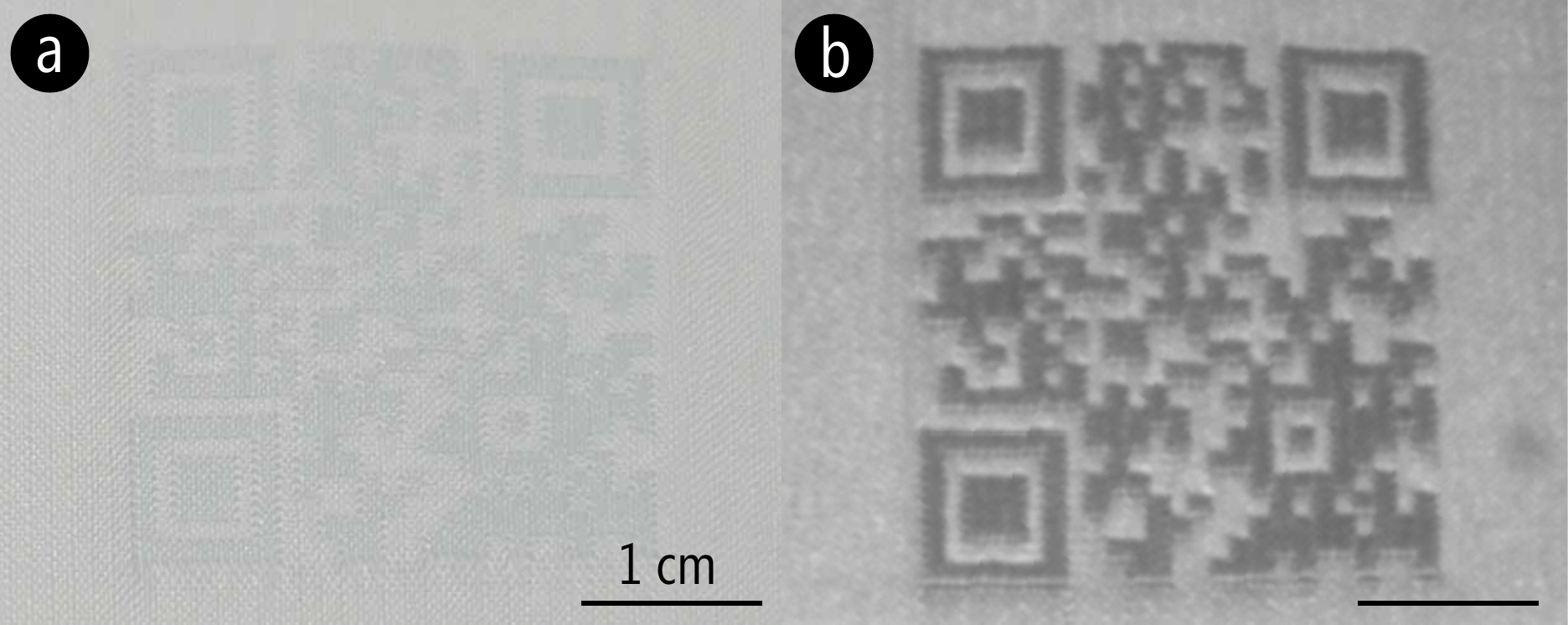}
  \caption{The baseline woven QR code without disguising treatment in the visible region (a) and in the NIR region (b).}
  \Description{Left: visible-light photo of a woven QR code on fabric without disguising strategies, where the pattern is faintly visible. Right: near-infrared image of a woven QR code with strong black-and-white contrast. }
  \label{fig:baseline}
\end{figure}

Although woven markers made with NIR-absorbing yarns appear relatively unobtrusive, they can still remain slightly perceptible under ambient light (\autoref{fig:baseline}).

Even the marker woven with only one type of yarn is visible under close inspection due to subtle surface unevenness caused by the layer-switching structure.
To further reduce this residual visibility while preserving NIR readability, we explored five disguising strategies: (a) yarn plying, (b) pattern decoying, (c) inkjet overprinting, (d) post-dyeing, and (e) dark-color blending (\autoref{fig:disguise}).
These strategies were selected for their practicality, as each is manufacturable across scales, from hand weaving to industrial production.
We evaluate all five strategies in \autoref{sec:evaluation}.

\paragraph{Yarn plying.}
We applied yarn plying by wrapping an additional yarn around the NIR-absorbing yarn (navy and white yarns in \autoref{fig:disguise}a). This outer yarn layer helps blend the marker into the surrounding textile by reducing local texture contrast on the surface.

\paragraph{Pattern decoying.}
We introduced surrounding ``decoy'' patterns that are visually similar to the genuine marker but not machine-readable. We hypothesized that these decoys make the overall textile pattern appear more spatially uniform, thereby reducing the perceptual salience of the actual marker compared to placing a single marker on an otherwise plain textile.

\paragraph{Inkjet overprinting.}
After weaving the marker, we added an additional color layer using a textile inkjet printer (Shima Seiki SIP-160F2L). 

\paragraph{Post-dyeing.}
We dyed the entire textile using a dyeing machine (Texam MC-12EL) to reduce the slight visible color difference between the yarns and create a more visually uniform appearance.

\paragraph{Dark-color blending.}
For dark textiles, normal black yarn and Teijin Heat Energy yarn, which is black under both visible and NIR regions, are paired as NIR-transparent and NIR-absorbing wefts to visually blend the marker into the surrounding textile. Dark textiles reduce visibility of markers by lowering shadow contrast due to surface unevenness.

\begin{figure}[bth]
    \centering
    \includegraphics[width=\linewidth]{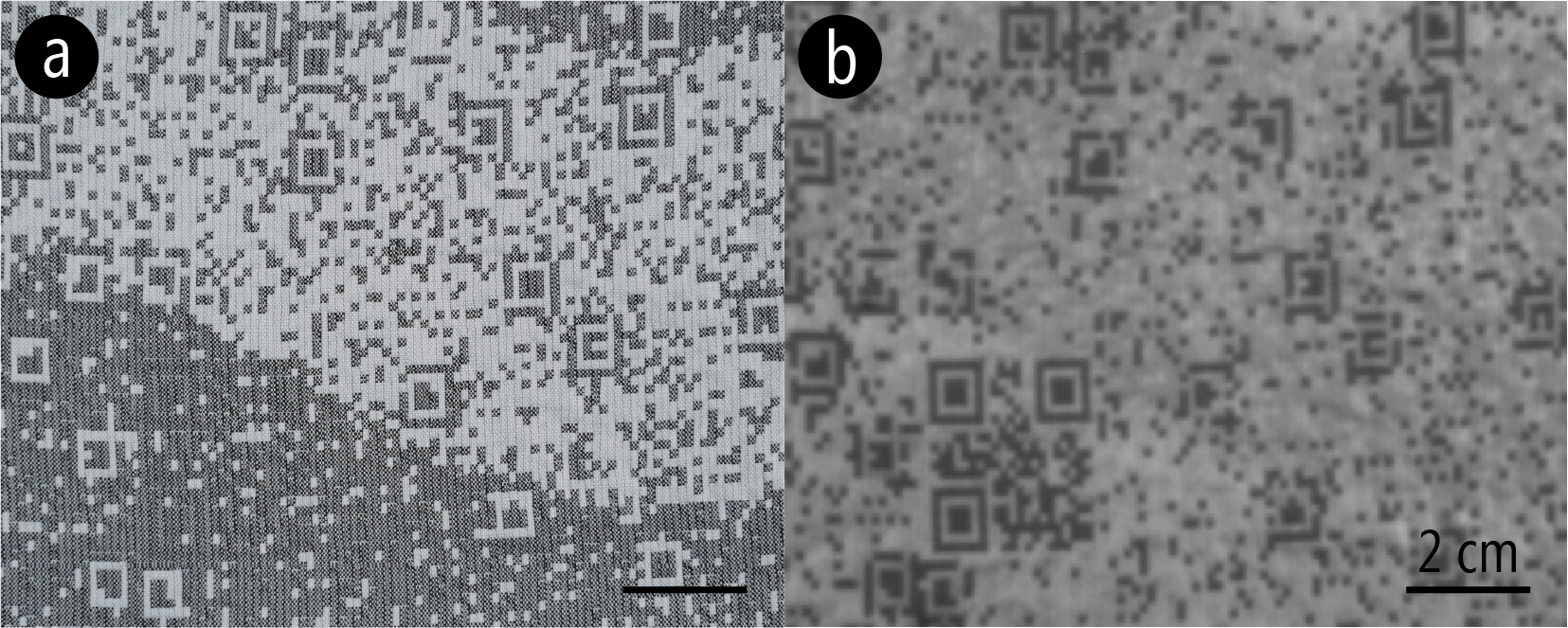}
    \caption{An example combining pattern decoying and dark-color blending across a large woven textile in the visible region (a) and in the NIR region (b).}
    \Description{Two side-by-side photos of a large textile densely tiled with many small woven squares. (a) Under visible light, the markers remain visually unobtrusive and blend into the fabric surface. (b) Under NIR imaging, the QR code appears with high contrast and remains machine-readable.}
    \label{fig:wave}
\end{figure}

\paragraph{Combining strategies.}
These strategies are not mutually exclusive and can be combined within a single textile to further reduce residual visibility while preserving NIR readability. \autoref{fig:wave} shows an example that combines multiple disguising strategies (pattern decoying and dark-color blending) across a large woven area with waved design. Under ambient light (\autoref{fig:wave}a), the markers remain visually unobtrusive within the surrounding textile, whereas under NIR imaging (\autoref{fig:wave}b), they stay clearly machine-readable.

\begin{figure*}[tb]
  \centering
  \includegraphics[width=1\linewidth]{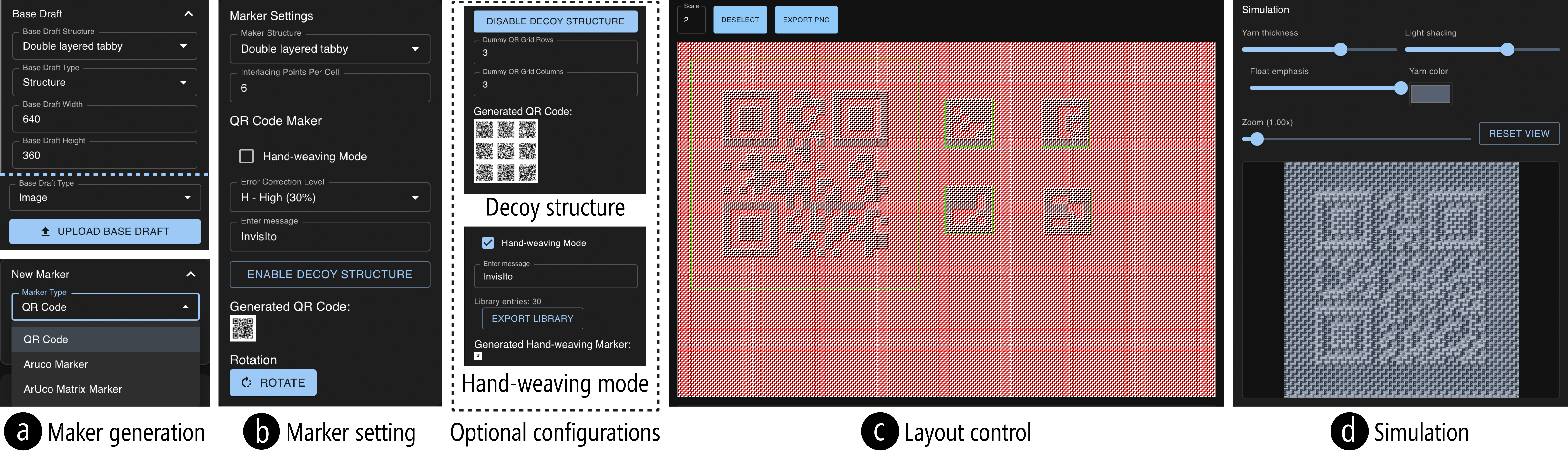}
  \caption{Workflow of the InvisIto design tool: (a) marker generation, (b) marker setting with optional configurations, (c) layout control, and (d) simulation. }
  \Description{Interface of the InvisIto design tool, with panels for base draft setup, marker generation, optional configuration, layout control, and simulation preview.}
  \label{fig:invisito-design-tool}
\end{figure*}

\subsection{Design Tool}
We developed a web-based design tool in TypeScript\footnote{\url{https://www.typescriptlang.org/}}, built on AdaCAD~\cite{devendorf_adacad_2023}, that supports marker authoring and integration within weave drafts.
The tool extends AdaCAD’s operator libraries and visualization environment with modules for marker generation, marker-specific parameterization, layout control, and woven-fabric simulation.
The design tool supports two types of markers: QR codes for digital data encoding and ArUco markers for binary input sensing and deformation tracking. 

As shown in \autoref{fig:invisito-design-tool}, the system integrates four core components into a cohesive workflow: \textit{marker generation} for encoding or tracking elements, \textit{marker settings} for configuring marker-specific parameters, \textit{layout control} for positioning markers within the textile design, and \textit{fabric simulation} for previewing woven appearance.
Users first define a base draft by selecting a built-in weave structure or importing an existing draft image.
They then generate a marker, adjust marker-specific parameters, and place the resulting marker within the textile layout.
The tool provides immediate visual feedback through both the generated weaving draft and a woven-fabric simulation view.

\paragraph{Marker generation.} 
Users begin by importing an existing weaving draft image (e.g., a \texttt{.png} file) or selecting a base structure type from a built-in library, including satin, twill, and plain weave.
The imported design defines a base canvas for marker placement. 
Users then select a marker type for subsequent configuration and placement.

\paragraph{Marker setting.} 
Depending on the selected marker type, the system generates:  

  \begin{itemize}[leftmargin=0.6cm]
    \item \emph{QR codes:} Users can embed a fixed message, such as a URL or text string.
    The system computes the minimum QR code resolution and pattern pixel size required for the given payload, based on the empirical thresholds reported in \autoref{sec:evaluation}, to support reliable detection after fabrication.
    Users can also specify the error-correction level and optionally enable additional configurations, including pattern decoying and hand-weaving mode.
    For pattern decoying, the system surrounds the true marker with a user-configurable dummy QR grid as a disguising method.
    In hand-weaving mode, textual payloads are converted into compact woven markers better suited to manual fabrication.
    The current implementation represents each marker as a deterministic small binary pattern composed of fixed \numproduct{2 x 2} binary blocks, followed by a quiet zone and optional rotation.
    Each generated marker is stored in a local library together with its content and marker ID, enabling later retrieval, reuse, and decoding by external tools.
    \item \emph{ArUco markers:} The tool supports both standalone ArUco markers and ArUco marker matrices.
    Standalone ArUco markers are intended for binary input sensing, while ArUco marker matrices support deformation tracking by distributing multiple uniquely identified markers across a configurable grid.
    The current implementation uses a \numproduct{4 x 4} ArUco dictionary.
    For standalone ArUco markers, users can set the marker ID.
    For ArUco matrices, users can specify matrix dimensions and spacing-related parameters for the generated layout.
  \end{itemize}

\paragraph{Layout control.} 
After generation, markers can be positioned within the draft through our layout control. 
This allows users to blend markers with the underlying weave structure and compose them within the textile layout before export. 
For QR codes and standalone ArUco markers, users can translate, rotate, and scale the marker to fit the weave structure or garment boundaries. 
For ArUco matrices, only global repositioning of the full marker set is supported. 
After each adjustment, the system automatically blends the marker with the underlying canvas structure. 

\paragraph{Simulation.} 
To support visual inspection prior to fabrication, the tool provides a woven fabric simulation view that renders the generated draft as an approximate textile surface. 
This preview aims to help the user assess how the marker is likely to appear once woven, rather than only as a binary draft image. 
The simulation interface includes controls for yarn appearance, including yarn thickness, light shading, float emphasis, yarn color, and zoom control, allowing users to inspect marker visibility under different visual conditions.

\subsection{Fabrication}
Since InvisIto relies only on specialized yarns, it does not introduce additional fabrication steps into conventional weaving workflows. The design tool automatically generates weaving drafts in standard weave notation, allowing direct interpretation by both manual and computer-controlled looms with minimal modification. This compatibility enables InvisIto to integrate into existing textile production pipelines. The method extends beyond rapid prototyping and can scale across diverse fabrication contexts.
In our implementation, lab-scale samples were woven on an Orika\footnote{\url{https://www.4dbox.com/lineup/orika}} customized Jacquard loom (Toyoshima Business System\footnote{\url{https://toyoshimabs.co.jp/}}), production-scale samples were commissioned at an external weaving mill, and handwoven samples were produced manually to demonstrate the method’s applicability across weaving contexts.
\autoref{tab:repro} summarizes the materials, loom settings, and fabrication parameters used in these implementations.

\begin{table}[th]
  \centering
  \caption{Summary of materials and methods.}
  \label{tab:repro}
  {\footnotesize
  \begin{tabularx}{\columnwidth}{p{0.3\columnwidth}X}
    \toprule
    \textbf{Category} & \textbf{Specification} \\
    \midrule
    \textbf{NIR-absorbing yarn} &
    SOLAMENT MKT2205A (\qty{150}{D}) and Teijin Heat Energy TWYCKU160 BCE167T48 (\qty{150}{D}) \\
    \textbf{Normal yarn} &
    Polyester yarns: \qty{75}{D} (warp) and \qty{150}{D} (weft). \\
    \textbf{Draft tool} &
    AdaCAD. \\
    \textbf{Loom} &
    Customized Jacquard loom Orika. Larger samples were manufactured at an external weaving mill. \\
    \textbf{NIR camera} & Omron Sentech
    STC-SBS34U3V-SWIRU \\
    \textbf{Spectral band} &
    Sensor bandwidth: \qtyrange[range-phrase={--}]{400}{1700}{\nano\meter}. \\
    \textbf{Imaging setup} &
    Camera-fabric distance: \qty{30}{\centi\meter}.
    View: top-down.
    Field of view: H\qty{17.1}{\degree}, V\qty{12.8}{\degree}.
    Mounting: tripod. \\
    \textbf{Lighting} &
    Two halogen lamps provide wide-band illumination. Ambient lighting: indoor. \\
    \textbf{Processing} & MacBook Pro (Apple M4, 24GB RAM) \\
    \bottomrule
  \end{tabularx}
  }
\end{table}

\subsection{Detection Pipeline}
\label{sec:detection}
After fabrication, InvisIto processes NIR image frames to interpret the woven markers. QR codes provide static information retrieval, while ArUco markers enable binary input sensing and real-time deformation tracking.

Our benchtop setup combines two halogen lamps, used as broadband NIR illumination, with an STC-SBS34U3V-SWIRU NIR camera from Omron Sentech\footnote{\url{https://sentech.co.jp/en}}, which has a spectral range of \qtyrange{400}{1700}{\nano\meter}.
In a more practical situation, the markers can also be decoded under ordinary ambient lighting using an affordable smartphone equipped with an NIR camera module (e.g., OUKITEL WP35 S\footnote{\url{https://oukitel.com/pages/oukitel-wp35-s-specs}\label{foot:oukitel}}).
Under NIR illumination, the NIR-absorbing yarn appears darker than the surrounding yarns, revealing the woven marker pattern. The captured frames are then passed to task-specific recognition modules:

\paragraph{QR code decoding}
For QR codes, captured NIR images were directly fed into the Dynamsoft barcode decoder\footnote{\url{https://www.dynamsoft.com/barcode-reader/overview/}} for detection and decoding.

\paragraph{Binary input sensing}
For binary input sensing, standalone woven ArUco markers are placed at discrete interaction locations. Each marker represents a binary state based on whether it is visible or occluded in the camera view. 
For each frame, OpenCV\footnote{\url{https://opencv.org/}\label{foot:opencv}}'s ArUco detector is used to identify the visible marker IDs. 
The presence or absence of each marker is then mapped to the corresponding input event.

\paragraph{Deformation tracking algorithm}
For deformation tracking, we developed a dynamic tracking algorithm for woven ArUco marker matrices.
The matrix consists of \numproduct{5 x 5} ArUco markers, each measuring \qty{7.5}{\milli\meter} square, arranged in a grid pattern at intervals of approximately \qty{30}{\milli\meter}.
We used the \texttt{4x4\_50} dictionary included in OpenCV\footref{foot:opencv} for the ArUco markers.
Our deformation tracking is based on the detection part of Deformable Dot Cluster Marker~\cite{narita_dynamic_2017}.
Let $\mathbf{x}_i(t)$ be the center coordinates of the $i$-th ArUco marker tracked at time $t$.
In the update step of the algorithm, when the ratio of black pixels in a square region of interest (ROI) $r_i(t)$ is greater than pre-defined threshold $\tau_\mathrm{B}$, we consider an ArUco marker to exist in ROI, and update the new position $\mathbf{x}_i(t)$ to the centroid of that region.
In the recovery step of the algorithm, we detect ArUco markers across the entire image instead of the ID of dot clusters, and overwrite the coordinates of tracked markers $\mathbf{x}_i(t)$.
If we would like to map the texture following the detected ArUco markers (as demonstrated in \ref{subsec:cushion}), the given texture image was distorted based on the relative positional relationship of multiple markers arranged in a grid.

\section{Technical Evaluation}
\label{sec:evaluation}
To characterize the performance of woven markers produced by InvisIto, we evaluated the system from three perspectives. First, we assessed the disguising methods in terms of perceived visual unobtrusiveness under ambient light and machine readability under NIR imaging. Second, we evaluated the decoding robustness of woven QR markers across different resolutions, error correction levels, weave structures, and washing cycles. Third, we evaluated the ArUco tracking performance. As a baseline condition, we used a QR marker with SOLAMENT 8\% and a normal white yarn (\autoref{fig:baseline}) without any disguising treatment for all the experiments.

\begin{figure*}[t]
    \centering
    \includegraphics[width=\linewidth]{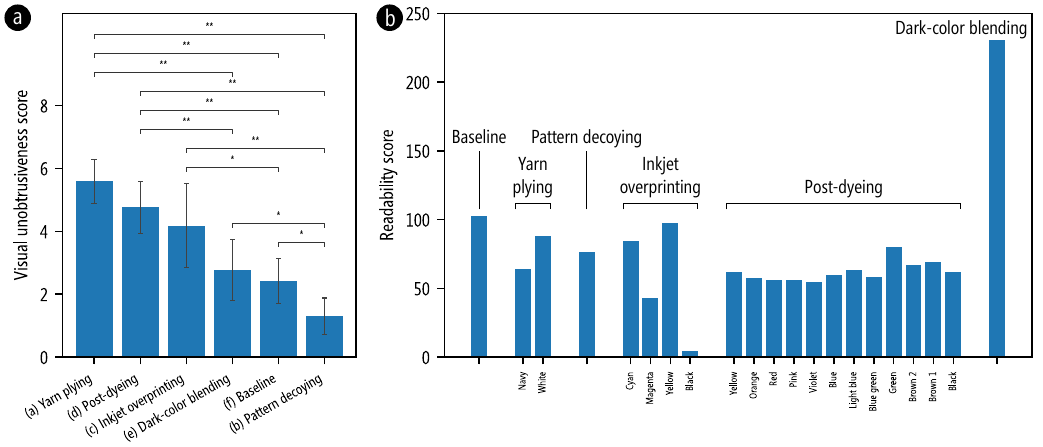}
    \caption{Evaluating visual unobtrusiveness and readability. a: The result of the user study shows perceived visual unobtrusiveness across the five disguising methods. Asterisks indicate statistical significance ($*$: $p < 0.05$ and $**$: $p < 0.01$). b: The between-class variance shows machine-readability of the QR code under NIR light.}
    \label{fig:eval-results}
\end{figure*}

\subsection{Visual Unobtrusiveness and Readability}
All QR codes share the same module layout, size (\qtyproduct{30.88 x 29.94}{\milli\meter}), and a high error-correction code; only the disguising treatment varies across conditions.

\paragraph{Visual unobtrusiveness under ambient light}
To evaluate the visual unobtrusiveness of woven markers, we conducted a user study with 17 participants (13 males, 4 females, age = 22.6 years old $\pm$ 1.4 \textit{s.d.}).
First, we instructed the participants that we are developing less obtrusive QR code markers.
They were given six samples of woven QR codes: (a) the code disguised with yarn plying (with a navy coiling yarn), (b) surrounded by decoys, (c) overlaid with light blue ink from an inkjet printer, (d) dyed with light blue dye, (e) blended with dark colors, as well as (f) the baseline QR code. They were asked to rank the disguised markers in order from the most noticeable (score 1) to the least noticeable (score 6). For each participant, we randomized the order of the six samples on the table. We calculated the mean value and the standard deviation of the score. As statistical tests, the Friedman test ($p < 0.01$) and pairwise tests were conducted.

Among the five disguising methods as shown in \autoref{fig:eval-results}a, (a) yarn plying, (d) post-dyeing, and (c) inkjet overprinting are significantly more unobtrusive than (f) baseline and (b) pattern decoying, respectively. On the other hand, there is no significant difference observed between (a), (d), and (c). Thus, we concluded that (a) yarn plying, (d) post-dyeing, and (c) inkjet overprinting are the methods that can better disguise the woven markers.

\paragraph{Readability under NIR light}
We also evaluated the machine-readability of the five disguising methods by quantifying the NIR contrast between the black and white regions in the woven code. For each disguising method, we captured NIR images and cropped each image to include only the QR code area. We then applied Otsu's thresholding method~\cite{otsu_threshold_1979} and computed the maximum between-class variance $\max \sigma^2_{\mathrm{B}}$ by averaging three samples, which was used as a readability score.

The result shown in \autoref{fig:eval-results}b revealed that most of the disguised conditions have readability comparable to the baseline. In particular, yarn plying and dyeing remain readable for all the colors tested while reducing visual obtrusiveness. As an exceptional case, dark-color blending has the highest readability, because only this condition used black wefts (i.e., Teijin Heat Energy and a normal black yarn), which have a higher contrast in the NIR region. On the other hand, the magenta and black inks of inkjet overprinting reduced contrast by more than $50\%$ of the baseline, resulting in substantially lower readability. We note that $\sigma^2_{\mathrm{B}}$ for the black QR code printed on white paper (i.e., a perfectly visible QR code) was \num{1109.14}, as a possible highest score.

To summarize the two experiments, dark-color blending is the most suitable option when a black textile is acceptable, and maximum readability is prioritized. Meanwhile, most of the disguising methods are effectively readable under NIR light, while reducing the visibility under ambient light with preferred color and design.


\subsection{QR Decoding Performance}
\label{sec:qr-eval}
We evaluated the decoding performance of our QR codes across four dimensions: module resolution, error correction code (ECC), weave structure, and washing durability.
A QR code is built from modules, the smallest black-and-white square units of the QR code matrix.
All samples used Version~1 QR codes (\numproduct{21 x 21} modules) encoded in byte mode.
Unless otherwise noted, each condition was tested over 1{,}000 decoding attempts at a fixed camera distance of \qty{30}{\centi\meter} under uniform NIR illumination.
We recorded both decoding success rate (\%) and time-to-decode (TtD).

\paragraph{Resolution threshold}
We first evaluated the minimum module resolution required for reliable decoding.
When a marker is woven, each QR module is formed from a small block of warp and weft yarns, so we express the module resolution in yarns per module, the number of warp and weft yarns spanning a single module.
We varied the module resolution from 0.5 to 4~yarns per module while keeping all the other imaging conditions fixed, using ECC--Q (2nd highest error correction level). 
We tested two marker samples that differed only in the color of the second weft yarn, using either white or gray, and both showed the same overall trend: decoding performance remained high at larger module sizes and dropped steeply once the resolution approached 3~yarns per module. \autoref{tab:resolution-threshold} indicates that the practical decoding threshold of the current system is approximately 3.5~yarns per module at a capture distance of \qty{30}{\centi\meter}.
TtD remained relatively stable across resolution levels for successful trials, averaging at \qty{26.4}{\milli\second}, suggesting that failures near the threshold were primarily caused by insufficient spatial sampling rather than increased computational cost.

\begin{table}[tbh]
    \centering
    \caption{QR decoding success rates (\%) across module resolutions for samples with white and gray second-weft yarns. }
    \label{tab:resolution-threshold}
    {\footnotesize
    \begin{tabular}{S[table-format=1.1]S[table-format=2.1]S[table-format=2.1]}
        \toprule
        {Module size [yarns/module]} & {White weft [\%]} & {Gray weft [\%]} \\
        \midrule
        4.0 & 97.1 & 96.5 \\
        3.5 & 89.6 & 86.5 \\
        3.0 & 46.6 & 36.2 \\
        2.5 & 28.3 & 25.4 \\
        \bottomrule
    \end{tabular}
    }
\end{table}

\paragraph{Decoding performance and latency of ECC}
We next evaluated how QR ECC affects decoding performance and latency near the resolution threshold. Across plain-weave samples with module resolution from 0.5 to 4~yarns per module, decoding strongly depended on ECC redundancy. ECC--H (highest error correction level) consistently performed best, achieving 91.9\% decoding success at 3~yarns per module, while ECC--L (low) and ECC--M (middle) both fell below 10\%. At 2.5~yarns per module, ECC--H still retained 44.7\% success, whereas lower-redundancy settings showed substantial degradation. At larger module resolution ($\geq$3.5~yarns per module), all ECC levels performed similarly, generally exceeding 85\% success. Latency results showed a similar trend: ECC--H remained below \qty{32.9}{\milli\second} near the threshold, while ECC--L and ECC--M often exceeded \qty{120}{\milli\second} or failed to decode. These results suggest that higher-redundancy settings such as ECC--Q and ECC--H are better suited to textile-integrated QR codes when marker size is constrained.

\paragraph{Effect of weave structure}
Because InvisIto embeds markers directly into the woven structure, we also examined how weave structure affects decoding reliability. We compared the three double-layered weave structures supported in our design tool: plain (or tabby, the structure we used in the other part of our paper), twill, and satin. Using ECC--H markers at 4~yarns per module, all three weave structures achieved similar decoding success, with plain weave at 93.6\%, twill at 91.9\%, and satin at 92.5\%.

\paragraph{Washing durability}
Finally, we evaluated the durability of woven QR markers under repeated laundering. We tested ECC--H plain-weave markers at 4~yarns per module after 0, 5, 10, and 20 washing cycles. The laundering procedure approximately followed ISO 6330, using a top-loading agitator washing machine and line drying. Decoding success remained high across all conditions. 
After 10 and 20 washing cycles, decoding success decreased modestly to 95.0\% and 87.4\%, respectively. Based on visual inspection of the captured NIR images, this decrease appeared more closely related to minor wrinkling and local weave distortion than to any substantial loss of marker contrast. Overall, the markers retained strong readability after repeated washing, suggesting that InvisIto can support garment-scale use over time.


\subsection{ArUco Tracking Performance}
\paragraph{Frame rates}
Next, we evaluated the sustainable frame rate required to detect and track ArUco markers. For three conditions where the ArUco marker array is stable, rotated, and distorted, 5,000 frames of \numproduct{1920 x 1080}~px grayscale images were captured from \qty{30}{\centi\meter} distance.
We then calculated the mean, standard deviation, 50th percentile (P50), 99th percentile (P99), and maximum computation time per frame (\autoref{tab:process-time}). We also calculated the 99th percentile frame rate. Note that we defined the completion time of detection or tracking when each frame finds at least one marker.

\begin{table}[tbh]
    \centering
    \caption{The computation time (\unit[mode=text,reset-text-family=false,reset-text-series=false,reset-text-shape=false]{\milli\second}) per frame by each thread and the 99th percentile frame rate.}
    \label{tab:process-time}
    {\footnotesize
    \begin{tabular}{llS[table-format=1.2(1),uncertainty-mode=separate]S[table-format=1.2]S[table-format=1.2]S[table-format=2.2]S[table-format=3.2]}
        \toprule
        Moves & Threads & {Mean $\pm$ \textit{s.d.}} & {P50} & {P99} & {Max} & {P99 FPS [\unit{\per\second}]} \\
        \midrule
        Stable & Detection & 5.13 +- 0.14 & 5.11 & 5.61 & 9.84 & 178.10 \\
        Stable & Tracking & 5.95 +- 0.21 & 5.93 & 6.57 & 13.92 & 152.16 \\
        Rotated & Detection & 5.43 +- 0.15 & 5.42 & 5.95 & 7.80 & 167.96 \\
        Rotated & Tracking & 6.25 +- 0.18 & 6.23 & 6.89 & 8.34 & 145.17 \\
        Distorted & Detection & 5.47 +- 0.39 & 5.42 & 6.71 & 10.18 & 148.94 \\
        Distorted & Tracking & 6.36 +- 1.05 & 6.25 & 8.00 & 31.32 & 125.01 \\
        \bottomrule
    \end{tabular}
    }
\end{table}

The result shown in \autoref{tab:process-time} indicates that 99\% of the frames are processed at 125~fps or higher.
This demonstrates that ArUco tracking is sufficiently possible in real time with our current setup.

\paragraph{Accuracy}

To evaluate tracking accuracy, we measured the median positional error of the tracked ArUco marker against a per-video ground truth, together with the tracking rate.
We used \numproduct{1920 x 1080}~px, 30~fps grayscale videos in which a deformed \(5 \times 5\) ArUco marker (\qty{24.5}{px} = \qty{7.3}{\milli\meter}) array moved on a conveyor belt at three distinct constant speeds (\qtylist{192.48;144.99;44.62}{\milli\meter\per\second}).
Each measurement was repeated three times and averaged.
To decide the ground-truth position for every frame, we derived a constant-velocity translation model from ArUco decodes.

\begin{table}[tbh]
    \centering
    \caption{Tracking rates and median positional errors.}\label{tab:tracking-rates}
    \footnotesize
    \begin{tabular}{S[table-format=3.2]S[table-format=1.2]S[table-format=2.1]S[table-format=1.2]}
        \toprule
        {\makecell{Velocity\\{[\unit{\milli\meter\per\second}]}}} &
        {\makecell{Movement/frame\\{[\unit{\milli\meter}]}}} &
        {\makecell{Tracking rate\\{[\unit{\percent}]}}} &
        {\makecell{Positional error\\{[\unit{\milli\meter}]}}} \\
        \midrule
        44.62 & 1.49 & 97.4 & 0.86 \\
        144.99 & 4.83 & 92.0 & 4.26 \\
        192.48 & 6.42 & 91.5 & 8.52 \\
        \bottomrule
    \end{tabular}
\end{table}

As shown in \autoref{tab:tracking-rates}, the tracking rates were consistently high (>91\%) across all conditions, while the positional error increased with speed, from \qtyrange[range-phrase={ to }]{0.86}{8.52}{\milli\meter}.
The accuracy decrease mainly came from the motion blur and the camera frame rate, which was more limited than what the DDCM~\cite{narita_dynamic_2017} algorithm expects.
Notably, the error grew disproportionately faster between the mid and highest speed conditions than between the lowest and mid ones, suggesting that the practical accuracy limit of our 30~fps setup lies near \qty{144.99}{\milli\meter\per\second}.


\section{Applications}
\label{sec:applications}
To examine the compatibility of InvisIto with different textile workflows, we explore the design space of NIR weaving across multiple functions and weaving contexts. These applications span information encoding, binary input sensing, and deformation tracking. They also demonstrate how NIR markers can be incorporated through fabrication methods across scales ranging from hand weaving to industrial Jacquard production.

\subsection{T-shirt with Anti-Counterfeiting Marker}
\begin{figure}[ht]
  \centering
  \includegraphics[width=1\linewidth]{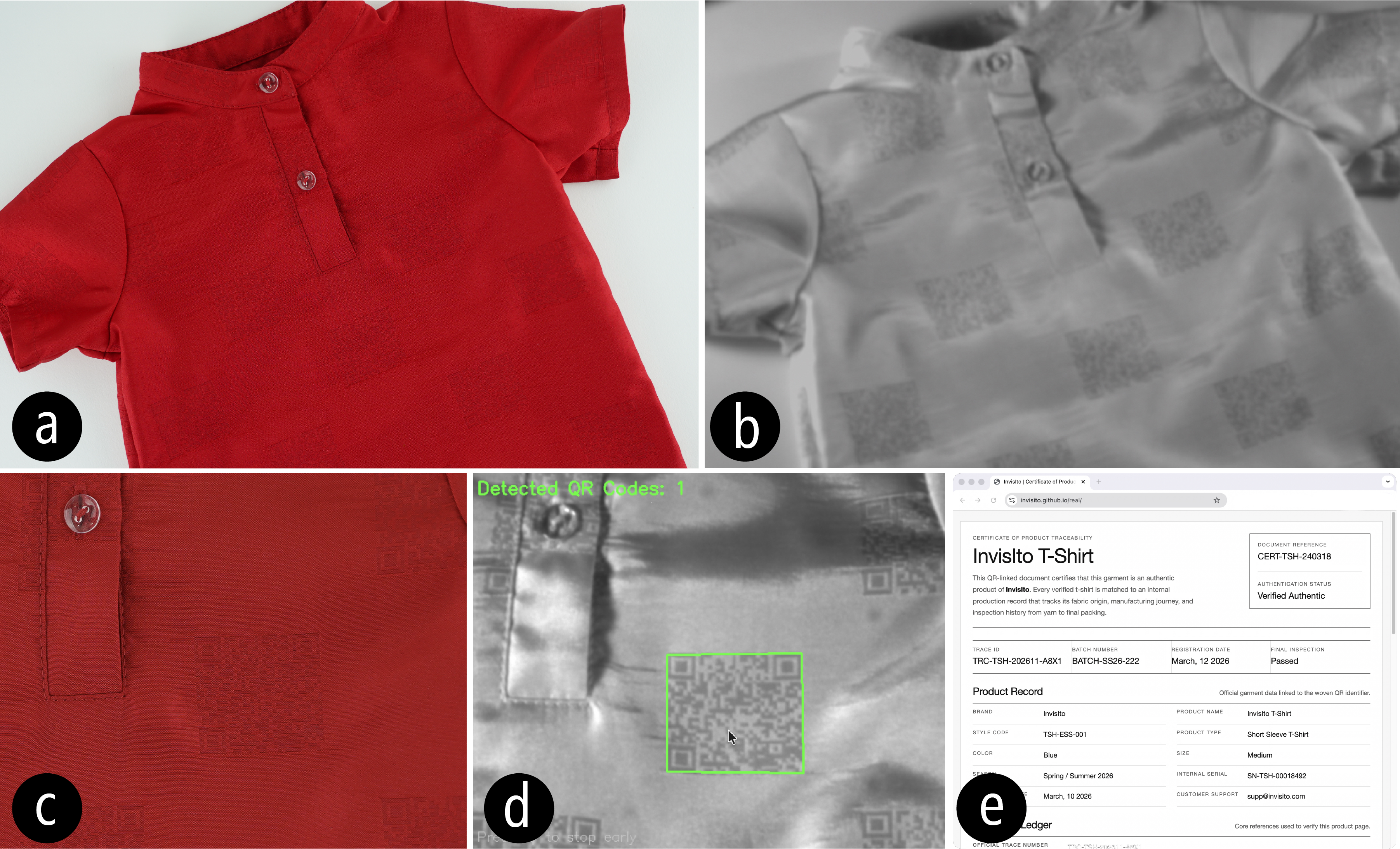}
  \caption{Anti-counterfeiting T-shirt.}
  \Description{Five-part figure showing a T-shirt with woven QR codes for authenticity verification. The first image shows the full garment under visible light. The second shows the same garment under near-infrared imaging, where multiple QR markers become visible. The third is a close-up of the fabric under visible light, where the markers are faint. The fourth shows NIR-based QR detection on the garment. The fifth shows the linked product verification webpage opened after scanning.}
  \label{fig:tshirt}
\end{figure}

InvisIto supports embedding authenticity information directly into the textile itself, removing the need for visible labels or attached tags. Unlike printed or attached authentication elements, the woven markers are structurally integrated into the fabric, making them inseparable from the garment while preserving its visual appearance. This offers a more unobtrusive and potentially more durable approach to authenticity marking.
We made a T-shirt containing multiple woven QR codes distributed across the garment for anti-counterfeiting (\autoref{fig:tshirt}a,c). To further reduce their visibility, the woven fabric was post-dyed. Every QR code encodes brand and serial information and links to a verification webpage\footnote{\url{https://invisito.github.io/real/}} (\autoref{fig:tshirt}e). The code remains visually less obtrusive in ambient light yet is reliably read by an NIR camera (\autoref{fig:tshirt}b,d).
By embedding verification data into the fabric structure itself, InvisIto turns the garment into a form of ``textile DNA,'' where machine-readable identity becomes an intrinsic part of the textile rather than an added label. 
This suggests potential uses beyond authenticity verification, such as supply chain tracing, provenance tracking, and garment life-cycle documentation.

\subsection{Handwoven Memory Wristband}
\begin{figure}[ht]
  \centering
  \includegraphics[width=1\linewidth]{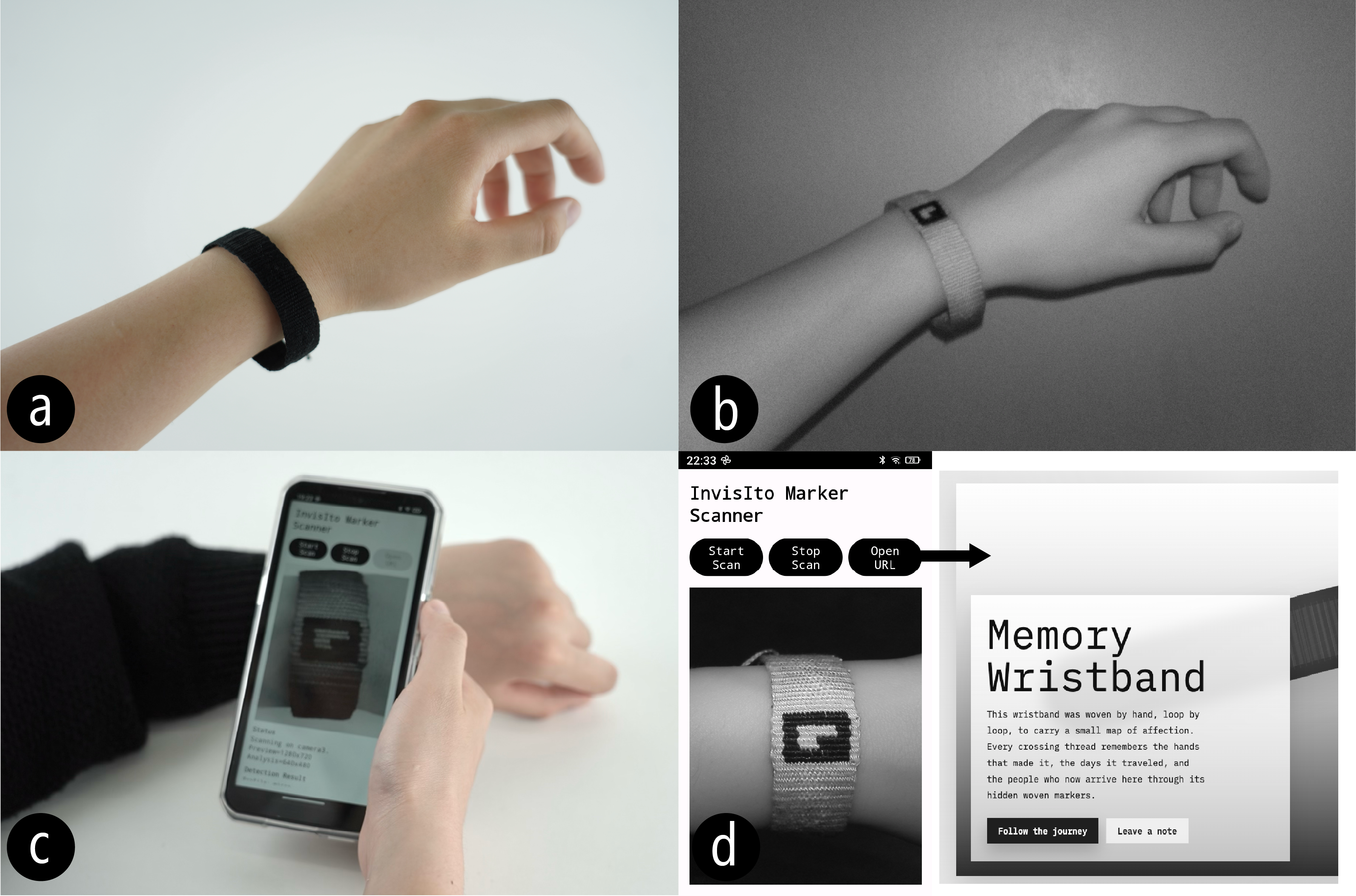}
  \caption{Handwoven memory wristband with an embedded marker.}
  \Description{A four-part figure showing a handwoven wristband application of InvisIto. The upper left shows the wristband worn under visible light, where the marker is not apparent. The upper right shows the same wristband under near-infrared imaging, where the embedded marker becomes visible. The lower left shows a user scanning the wristband with a smartphone. The lower right shows the scanning interface and a linked webpage titled Memory Wristband.}
  \label{fig:wristband}
\end{figure}

We created a wristband as a bespoke case study to show how InvisIto can support small-scale, highly personalized textile production (\autoref{fig:wristband}a). 
The piece explores fabric as a memory-bearing medium, drawing on prior work showing that clothing and textile making are often closely tied to identity, care, memory, and continuity with the past ~\cite{Riley2008, BuseTwigg2016}. 
In many familial and craft contexts, handmade textiles are exchanged, kept, and passed down as intimate objects of care, allowing them to accumulate personal and intergenerational significance over time.
Building on this idea, the weaver embedded unobtrusive markers directly into the wristband’s woven structure and linked them to a hidden URL accessible via NIR detection (\autoref{fig:wristband}b).
Scanning the visually unobtrusive marker with our mobile phone application on an OUKITEL WP35 S\footref{foot:oukitel}
provides direct access to the linked webpage (\autoref{fig:wristband}c). The device includes a built-in infrared camera for marker scanning. Through this webpage, users can leave comments and add personal content, allowing the wristband’s journey and the interactions it gathers over time to be recorded (\autoref{fig:wristband}d).
Because the markers are woven into the textile itself rather than added later as external tags or labels, they preserve the piece’s look and feel while introducing a machine-readable digital layer. 
The wristband thus becomes a digital-physical keepsake that can accumulate personal meaning over time.

\subsection{Smart Furniture Upholstery}
\begin{figure}[!ht]
  \centering
  \includegraphics[width=1\linewidth]{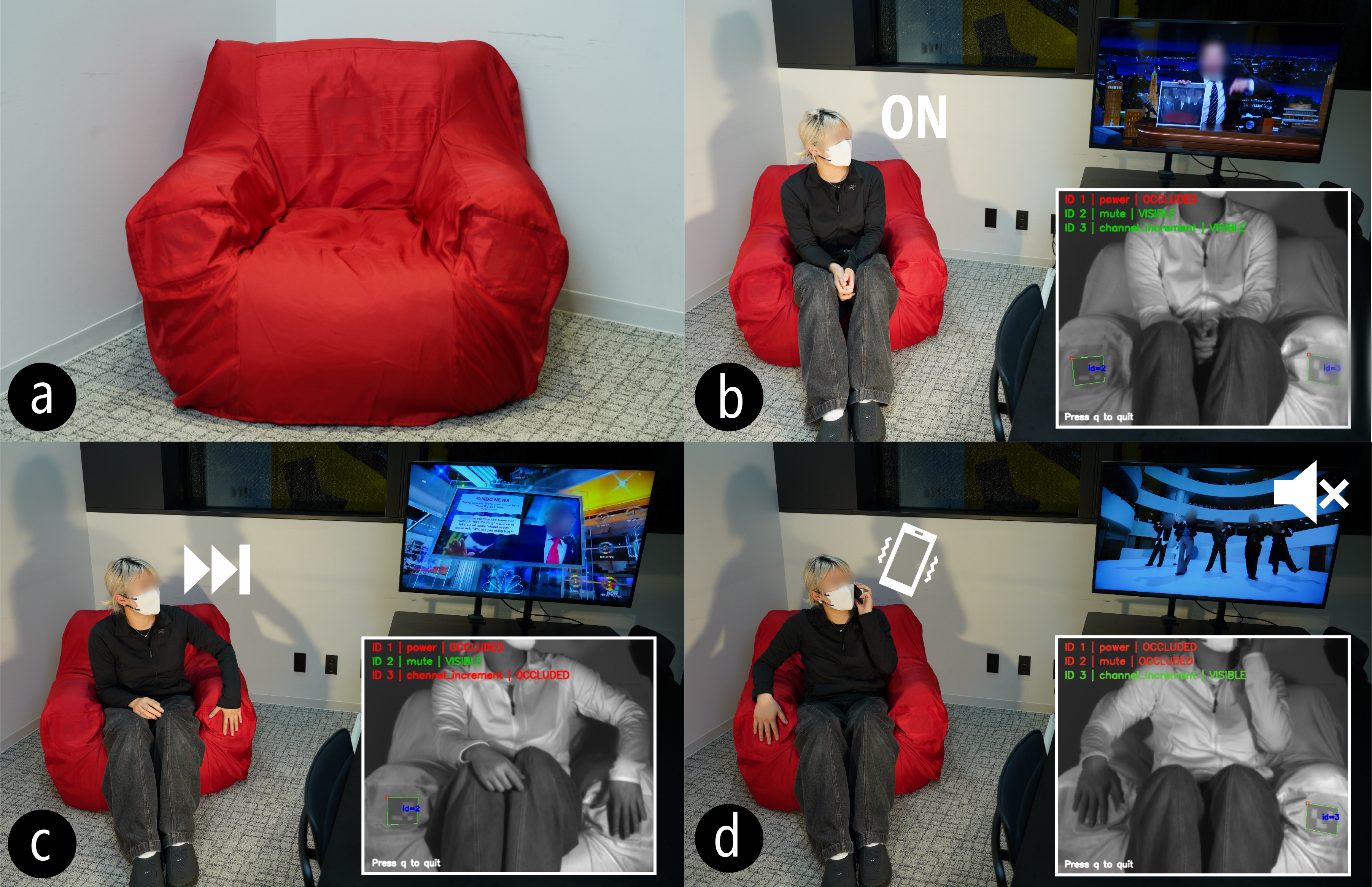}
  \caption{Sofa-integrated TV control using industrially woven markers.}
  \Description{Four-part figure showing a sofa upholstery application using InvisIto. The first panel shows a red sofa with no visible controls or markers on its surface. The remaining three panels show a seated user interacting with the sofa to control a television through different occlusion gestures. One panel shows the power function, one shows channel change, and one shows mute. Each interaction panel includes an inset near-infrared camera view showing the tracked marker regions and whether each marker is visible or occluded.}
  \label{fig:sofa}
\end{figure}

To demonstrate InvisIto's readiness for industrial-scale deployment, we fabricated sofa upholstery with ArUco markers woven into the fabric at an industrial weaving mill (\autoref{fig:sofa}a). 
A mounted camera feeds the live view into our detection pipeline, where each marker is interpreted as a discrete binary state based on whether it is visible or occluded. This enables button-like interactions without requiring contact localization or gesture recognition. We assigned three markers to different control functions: a power toggle on the seat back (\autoref{fig:sofa}b), channel increment on the left armrest (\autoref{fig:sofa}c), and mute on the right armrest (\autoref{fig:sofa}d). 
The resulting upholstery acts as a passive, zero-power textile interface remaining visually similar to standard upholstery, while sensing and processing are provided by the external camera system.
%

\subsection{Deformation-Aware AR Cushion}\label{subsec:cushion}
\begin{figure}[ht]
  \centering
  \includegraphics[width=1\linewidth]{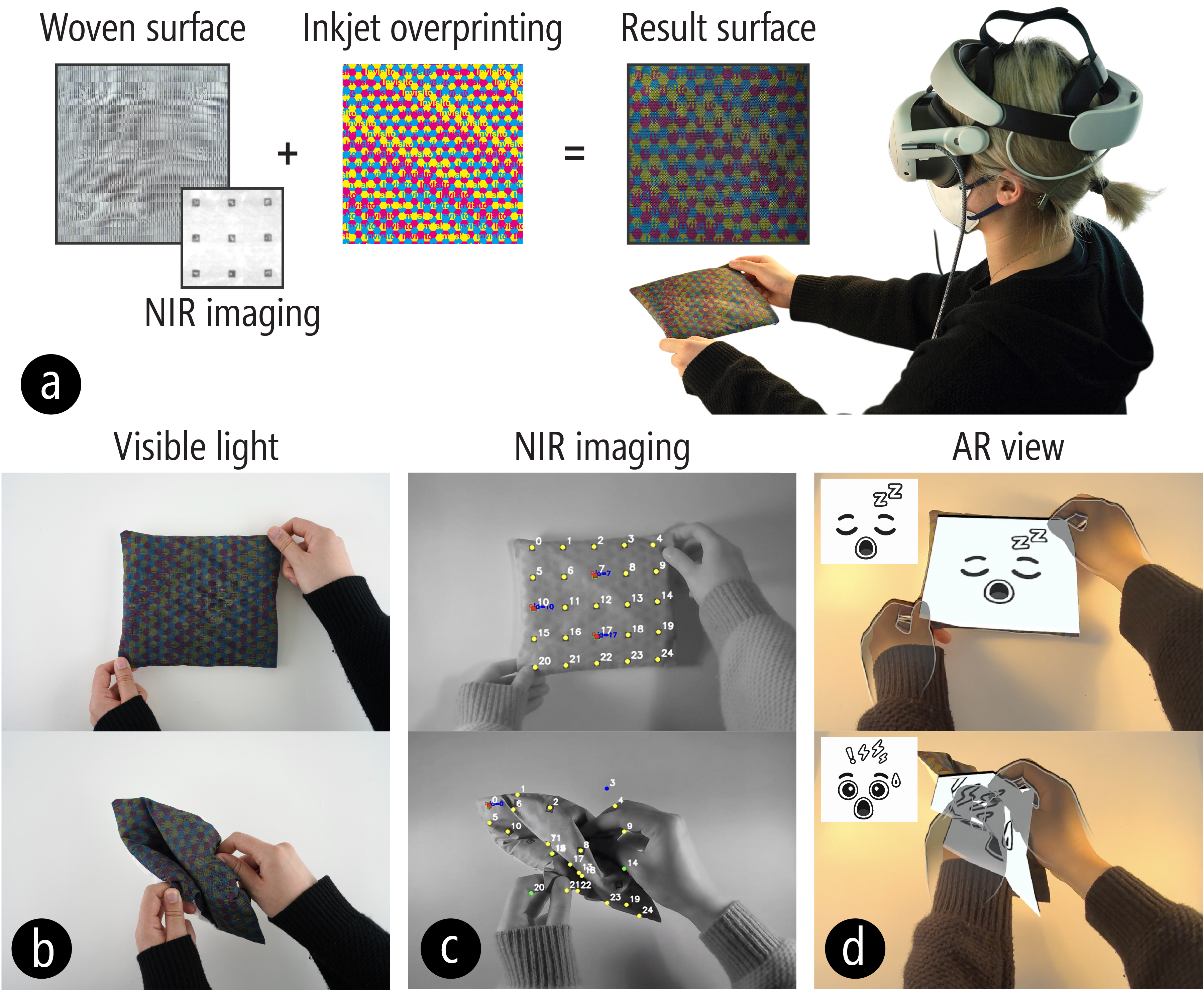}
  \caption{Deformation-aware AR interaction on a woven cushion with embedded markers.}
  \Description{Four-part figure showing an AR cushion application using InvisIto. The top panel illustrates the surface construction process: a woven surface with embedded markers is combined with a colorful inkjet overprint to produce a final textile appearance, alongside a person viewing the cushion through a head-mounted display. The lower panel shows: (a) the cushion under visible light in flat and deformed states, (b) near-infrared views of the same states with tracked marker points overlaid, (c) corresponding augmented reality views, where different virtual facial expressions are displayed in response to the cushion's deformation.}
  \label{fig:cushion}
\end{figure}

We created an augmented reality (AR) cushion with InvisIto’s ArUco matrix markers embedded into its top surface (\autoref{fig:cushion}a). To visually conceal the marker matrix, we applied inkjet overprinting, making it less perceptible under normal viewing conditions. The piece serves as a low-cost sensing surface by capturing geometric changes in the woven markers under interaction. In the head-mounted display (HMD), our tracking pipeline continuously estimates the cushion’s deformation (\autoref{fig:cushion}b) by tracking the ArUco matrix as shown in (\autoref{fig:cushion}c). The AR texture rendered in the HMD view then updates in real time to follow the fabric’s physical deformation (\autoref{fig:cushion}d), enabling deformation-aware AR rendering that remains spatially registered to the textile throughout interaction.

\section{Limitations and Future Work}
\label{sec:discussion}
\subsection{Visibility of Woven Markers}
Woven markers by InvisIto are visually unobtrusive, not entirely invisible. They stay perceptible on close inspection, and their salience depends on the weave design and the disguising strategy. InvisIto therefore reduces the visual prominence of machine-readable markers but does not make them imperceptible. Future work should explore a broader range of weaving techniques and disguising strategies to further reduce marker salience.


\subsection{Resolution and Yarn Type}
Yarn type and thickness impose important constraints on the resolution of woven markers. 
Yarn thickness and warp/weft densities define the smallest possible unit in the woven marker. 
Using coarser yarns or lower warp and weft densities results in larger woven pixels, requiring physically larger markers to encode the same payload and thereby reducing the achievable resolution. 
In practice, this makes it difficult to embed larger amounts of information without producing markers that are impractical in size. Material availability introduces an additional constraint. Our method currently supports only polyester fabrics because the commercially available IR-absorbing yarns used in our implementation are polyester-based.
Future work should therefore investigate finer yarns, higher-density weaving, and broader material compatibility to improve encoding capacity while extending InvisIto to a wider range of textile substrates.

\subsection{Beyond Conventional Fiducial Markers}
InvisIto encodes information using QR codes and ArUco markers, both of which were originally designed for printed, rigid surfaces with high contrast. We adopted these formats to maintain compatibility with existing computer vision pipelines. However, their underlying visual and geometric assumptions, including square modules, quiet zones, and rectilinear layouts, do not align well with the material properties of woven textiles. Future work could instead investigate encoding schemes that are native to textile structure and visual composition. Identifiers could be encoded through organic motifs, warp–weft relationships, weave density, or multi-layer structures~\cite{multilayer}. Such approaches could allow machine-readable information to become part of the textile's visual and structural design, rather than appearing as a separate graphical element.

\subsection{Fabric Deformation and Occlusion}
Fabric deformation remains a key challenge for woven marker decoding and tracking. Unlike rigid substrates, textiles stretch, wrinkle, and fold during use, which can distort marker geometry and reduce detection reliability. Occlusion also presents a particularly strong constraint because a marker covered by a hand or fabric fold cannot be detected. In our deformation tracking algorithm, however, the position of a temporarily undetected ArUco marker can be estimated from the surrounding marker grid, allowing the system to tolerate limited marker loss. 
Future work could extend this principle of spatial redundancy to information encoding by distributing repeating patterns, or textile-native organic motifs. Such encoding schemes could improve robustness to local deformation and partial occlusion while integrating machine-readable information more naturally into the visual and structural composition of textiles.

\section{Conclusion}
\label{sec:conclusion}
We introduced InvisIto, a method for weaving visually unobtrusive, machine-readable markers into textiles using NIR-absorbing yarns. 
InvisIto combines a design tool for embedding markers into weaving drafts, five disguising strategies that further reduce marker visibility under ambient light, and a detection pipeline for decoding and tracking under NIR imaging. 
With this approach, the system supports both woven QR codes for data encoding and woven ArUco markers for binary input and deformation tracking.
Through applications spanning hand weaving, Jacquard weaving, and industrial production, we validated that InvisIto can integrate with existing textile workflows to support weaving NIR markers across multiple fabrication contexts.

\begin{acks}
This research was supported by JSPS KAKENHI Grant Number JP25H01151, JST FOREST Program Grant Number JPMJFR232V, and JST ASPIRE Program Grant Number JPMJAP2401. We also would like to thank Rin Ishiguro and Yutaro Nakaya for their support on figure design, Rong Wu for assisting with the applications development, Ko Fujino for videography, and Mei Takamisawa for assisting with the evaluation.
\end{acks}

\bibliographystyle{ACM-Reference-Format}
\bibliography{references}

@inproceedings{multilayer,
author = {Frost, Devon and Jacobs, Jennifer},
title = {Warped Tour: Analyzing Design Tools Support for Multilayer Weaving Production in HCI},
year = {2026},
isbn = {9798400725630},
publisher = {Association for Computing Machinery},
address = {New York, NY, USA},
url = {https://doi.org/10.1145/3800645.3812988},
doi = {10.1145/3800645.3812988},
booktitle = {Proceedings of the 2026 Designing Interactive Systems Conference},
pages = {4501–4518},
numpages = {18},
location = {
},
series = {DIS '26}
}

@inproceedings{zebrasense,
author = {Wu, Tony and Fukuhara, Shiho and Gillian, Nicholas and Sundara-Rajan, Kishore and Poupyrev, Ivan},
title = {ZebraSense: A Double-sided Textile Touch Sensor for Smart Clothing},
year = {2020},
isbn = {9781450375146},
publisher = {Association for Computing Machinery},
address = {New York, NY, USA},
url = {https://doi.org/10.1145/3379337.3415886},
doi = {10.1145/3379337.3415886},
booktitle = {Proceedings of the 33rd Annual ACM Symposium on User Interface Software and Technology},
pages = {662–674},
numpages = {13},
location = {Virtual Event, USA},
series = {UIST '20}
}

@inproceedings{Xiao2025,
author = {Xiao, Chang},
title = {ReactFold: Towards Camera-based Tangible Interaction on Passive Paper Artifacts},
year = {2025},
isbn = {9798400711978},
publisher = {Association for Computing Machinery},
address = {New York, NY, USA},
url = {https://doi.org/10.1145/3689050.3704926},
doi = {10.1145/3689050.3704926},
booktitle = {Proceedings of the Nineteenth International Conference on Tangible, Embedded, and Embodied Interaction},
articleno = {29},
numpages = {12},
location = {
},
series = {TEI '25}
}

@inproceedings{Zheng2020,
author = {Zheng, Clement and Gyory, Peter and Do, Ellen Yi-Luen},
title = {Tangible Interfaces with Printed Paper Markers},
year = {2020},
isbn = {9781450369749},
publisher = {Association for Computing Machinery},
address = {New York, NY, USA},
url = {https://doi.org/10.1145/3357236.3395578},
doi = {10.1145/3357236.3395578},
booktitle = {Proceedings of the 2020 ACM Designing Interactive Systems Conference},
pages = {909–923},
numpages = {15},
location = {Eindhoven, Netherlands},
series = {DIS '20}
}

@article{barcode,
  author  = {Jeong, Sang-Mi and Yang, Jonguk and Pak, Jeong Hyeok and Seo, Keumyoung and Lim, Taekyung and Ju, Sanghyun},
  title   = {Real-Time Information-Variable Invisible Barcode Comprising Freely Deformable Infrared-Emitting Yarns},
  journal = {ACS Applied Materials \& Interfaces},
  year    = {2021},
  volume  = {13},
  number  = {34},
  pages   = {41046-41055},
  doi     = {10.1021/acsami.1c10052},
  note    = {PMID: 34402614},
  url     = {https://doi.org/10.1021/acsami.1c10052}
}

@article{Riley2008,
  author = {Jill Riley},
  title = {Weaving an enhanced sense of self and a collective sense of self through creative textile-making},
  journal = {Journal of Occupational Science},
  year = {2008},
  volume = {15},
  number = {2},
  pages = {63--73},
  doi = {10.1080/14427591.2008.9686611},
  url = {https://doi.org/10.1080/14427591.2008.9686611}
}

@article{BuseTwigg2016,
  author = {Christina E. Buse and Julia Twigg},
  title = {Materialising memories: exploring the stories of people with dementia through dress},
  journal = {Ageing \& Society},
  year = {2016},
  volume = {36},
  number = {6},
  pages = {1115--1135},
  doi = {10.1017/S0144686X15000185},
  url = {https://doi.org/10.1017/S0144686X15000185}
}

@inproceedings{sugiyama_weaving_2026,
  author    = {Sugiyama, Hal and Lee, Hsuanling and Fujino, Hanako and 
               Kuwana, Mayuka and Dogan, Mustafa Doga and He, Liang and 
               Narumi, Koya},
  title     = {Weaving and Disguising Infrared Markers toward Invisible Textile Interaction},
  booktitle = {Extended Abstracts of the 2026 CHI Conference on Human Factors 
               in Computing Systems},
  series    = {CHI EA '26},
  year      = {2026},
  pages     = {1--5},
  publisher = {Association for Computing Machinery},
  address   = {New York, NY, USA},
  location  = {Barcelona, Spain},
  doi       = {10.1145/3772363.3799013},
  isbn      = {979-8-4007-2281-3/2026/04},
}

@inproceedings{poupyrev_project_2016,
	address = {New York, NY, USA},
	series = {{CHI} '16},
	title = {Project {Jacquard}: {Interactive} {Digital} {Textiles} at {Scale}},
	isbn = {978-1-4503-3362-7},
	shorttitle = {Project {Jacquard}},
	url = {https://dl.acm.org/doi/10.1145/2858036.2858176},
	doi = {10.1145/2858036.2858176},
	urldate = {2025-09-09},
	booktitle = {Proceedings of the 2016 {CHI} {Conference} on {Human} {Factors} in {Computing} {Systems}},
	publisher = {Association for Computing Machinery},
	author = {Poupyrev, Ivan and Gong, Nan-Wei and Fukuhara, Shiho and Karagozler, Mustafa Emre and Schwesig, Carsten and Robinson, Karen E.},
	month = may,
	year = {2016},
	pages = {4216--4227},
}

@article{kuusk_crafting_nodate,
	title = {Crafting {Qualities} in {Designing} {QR}-coded {Embroidery} and {Bedtime} {Stories}},
	language = {en},
    year = {2014},
	author = {Kuusk, Kristi and Wensveen, Stephan and Tomico, Oscar},
    journal = {Proceedings of the Art of Research V Conference},
    month     = nov,
    address   = {Helsinki, Finland},
    pages     = {1--12},
}

@article{nofal_initiating_2020,
	title = {Initiating android phone technology using {QR} codes to make innovative functional clothes},
	volume = {32},
	copyright = {https://www.emerald.com/insight/site-policies},
	issn = {0955-6222},
	url = {http://www.emerald.com/ijcst/article/32/6/935-951/121625},
	doi = {10.1108/IJCST-12-2018-0153},
	language = {en},
	number = {6},
	urldate = {2025-09-09},
	journal = {International Journal of Clothing Science and Technology},
	author = {Nofal, Reem M.},
	month = apr,
	year = {2020},
	pages = {935--951},
}

@misc{halimi_garment_2022,
	title = {Garment {Avatars}: {Realistic} {Cloth} {Driving} using {Pattern} {Registration}},
	shorttitle = {Garment {Avatars}},
	url = {http://arxiv.org/abs/2206.03373},
	doi = {10.48550/arXiv.2206.03373},
	urldate = {2025-09-09},
	publisher = {arXiv},
	author = {Halimi, Oshri and Prada, Fabian and Stuyck, Tuur and Xiang, Donglai and Bagautdinov, Timur and Wen, He and Kimmel, Ron and Shiratori, Takaaki and Wu, Chenglei and Sheikh, Yaser},
	month = jun,
	year = {2022},
	note = {arXiv:2206.03373 [cs]},
}

@inproceedings{xu_bit_2025,
	address = {New York, NY, USA},
	series = {{CHI} '25},
	title = {{BIT}: {Battery}-free, {IC}-less and {Wireless} {Smart} {Textile} {Interface} and {Sensing} {System}},
	isbn = {979-8-4007-1394-1},
	shorttitle = {{BIT}},
	url = {https://dl.acm.org/doi/10.1145/3706598.3713100},
	doi = {10.1145/3706598.3713100},
	urldate = {2025-09-09},
	booktitle = {Proceedings of the 2025 {CHI} {Conference} on {Human} {Factors} in {Computing} {Systems}},
	publisher = {Association for Computing Machinery},
	author = {Xu, Weiye and Li, Tony and Wang, Yuntao and Yang, Xing-Dong and Wu, Te-Yen},
	month = apr,
	year = {2025},
	pages = {1--18},
}

@article{tone_fibar_2020,
	title = {{FibAR}: {Embedding} {Optical} {Fibers} in {3D} {Printed} {Objects} for {Active} {Markers} in {Dynamic} {Projection} {Mapping}},
	volume = {26},
	issn = {1077-2626, 1941-0506, 2160-9306},
	shorttitle = {{FibAR}},
	url = {http://arxiv.org/abs/2002.02159},
	doi = {10.1109/TVCG.2020.2973444},
	number = {5},
	urldate = {2025-09-09},
	journal = {IEEE Transactions on Visualization and Computer Graphics},
	author = {Tone, Daiki and Iwai, Daisuke and Hiura, Shinsaku and Sato, Kosuke},
	month = may,
	year = {2020},
	note = {arXiv:2002.02159 [cs]},
	pages = {2030--2040},
}

@article{jankowski-mihulowicz_identification_2023,
	title = {Identification {Efficiency} in {RFIDtex} {Enabled} {Washing} {Machine}},
	volume = {11},
	issn = {2169-3536},
	url = {https://ieeexplore.ieee.org/document/10256168},
	doi = {10.1109/ACCESS.2023.3317510},
	urldate = {2025-08-09},
	journal = {IEEE Access},
	author = {Jankowski-Mihułowicz, Piotr and Pawłowicz, Bartosz and Kołcz, Marek and Węglarski, Mariusz},
	year = {2023},
	pages = {103814--103829},
}

@incollection{iezzi_fiber_2024,
	title = {Fiber and {Fabric}-{Integrated} {Tracing} {Technologies} for {Textile} {Sorting} and {Recycling}},
	copyright = {© 2024 Scrivener Publishing LLC},
	isbn = {978-1-394-21429-7},
	url = {https://onlinelibrary.wiley.com/doi/abs/10.1002/9781394214297.ch18},
	language = {en},
	urldate = {2025-08-09},
	booktitle = {Technology {Innovation} for the {Circular} {Economy}},
	publisher = {John Wiley \& Sons, Ltd},
	author = {Iezzi, Brian and Shtein, Max and Wang, Tairan and Rothschild, Mordechai},
	year = {2024},
	doi = {10.1002/9781394214297.ch18},
	note = {Section: 18
\_eprint: https://onlinelibrary.wiley.com/doi/pdf/10.1002/9781394214297.ch18},
	pages = {223--237},
}

@article{iezzi_polymeric_2023,
	title = {Polymeric {Photonic} {Crystal} {Fibers} for {Textile} {Tracing} and {Sorting}},
	volume = {8},
	copyright = {© 2023 Wiley-VCH GmbH},
	issn = {2365-709X},
	url = {https://onlinelibrary.wiley.com/doi/abs/10.1002/admt.202201099},
	doi = {10.1002/admt.202201099},
	language = {en},
	number = {6},
	urldate = {2025-08-09},
	journal = {Advanced Materials Technologies},
	author = {Iezzi, Brian and Coon, Austin and Cantley, Lauren and Perkins, Bradford and Doran, Erin and Wang, Tairan and Rothschild, Mordechai and Shtein, Max},
	year = {2023},
	note = {\_eprint: https://advanced.onlinelibrary.wiley.com/doi/pdf/10.1002/admt.202201099},
	pages = {2201099},
}

@inproceedings{steimle_flexpad_2013,
	address = {New York, NY, USA},
	series = {{CHI} '13},
	title = {Flexpad: highly flexible bending interactions for projected handheld displays},
	isbn = {978-1-4503-1899-0},
	shorttitle = {Flexpad},
	url = {https://doi.org/10.1145/2470654.2470688},
	doi = {10.1145/2470654.2470688},
	urldate = {2025-08-07},
	booktitle = {Proceedings of the {SIGCHI} {Conference} on {Human} {Factors} in {Computing} {Systems}},
	publisher = {Association for Computing Machinery},
	author = {Steimle, Jürgen and Jordt, Andreas and Maes, Pattie},
	month = apr,
	year = {2013},
	pages = {237--246},
}

@inproceedings{punpongsanon_deforme_2013,
	address = {New York, NY, USA},
	series = {{SA} '13},
	title = {{DeforMe}: projection-based visualization of deformable surfaces using invisible textures},
	isbn = {978-1-4503-2632-2},
	shorttitle = {{DeforMe}},
	url = {https://doi.org/10.1145/2542284.2542292},
	doi = {10.1145/2542284.2542292},
	urldate = {2025-08-07},
	booktitle = {{SIGGRAPH} {Asia} 2013 {Emerging} {Technologies}},
	publisher = {Association for Computing Machinery},
	author = {Punpongsanon, Parinya and Iwai, Daisuke and Sato, Kosuke},
	month = nov,
	year = {2013},
	pages = {1--3},
}

@inproceedings{narita_dynamic_2015,
	address = {New York, NY, USA},
	series = {{VRST} '15},
	title = {Dynamic projection mapping onto a deformable object with occlusion based on high-speed tracking of dot marker array},
	isbn = {978-1-4503-3990-2},
	url = {https://doi.org/10.1145/2821592.2821618},
	doi = {10.1145/2821592.2821618},
	urldate = {2025-08-06},
	booktitle = {Proceedings of the 21st {ACM} {Symposium} on {Virtual} {Reality} {Software} and {Technology}},
	publisher = {Association for Computing Machinery},
	author = {Narita, Gaku and Watanabe, Yoshihiro and Ishikawa, Masatoshi},
	month = nov,
	year = {2015},
	pages = {149--152},
}

@article{maida_claycode_2025,
	title = {Claycode: {Stylable} and {Deformable} {2D} {Scannable} {Codes}},
	volume = {44},
	issn = {0730-0301, 1557-7368},
	shorttitle = {Claycode},
	url = {https://dl.acm.org/doi/10.1145/3730853},
	doi = {10.1145/3730853},
	language = {en},
	number = {4},
	urldate = {2025-08-07},
	journal = {ACM Transactions on Graphics},
	author = {Maida, Marco and Crescini, Alberto and Perronet, Marco and Camuffo, Elena},
	month = aug,
	year = {2025},
	pages = {1--14},
}

@inproceedings{chiu_fabric_2025,
	address = {New York, NY, USA},
	series = {{CHI} {EA} '25},
	title = {Fabric {Tag}\#: {Envisioning} a {UXD} {Solution} for {Textile} {Culture} {Preservation} and {Fashion} {Sustainability}},
	isbn = {979-8-4007-1395-8},
	shorttitle = {Fabric {Tag}\#},
	url = {https://doi.org/10.1145/3706599.3720302},
	doi = {10.1145/3706599.3720302},
	urldate = {2025-08-06},
	booktitle = {Proceedings of the {Extended} {Abstracts} of the {CHI} {Conference} on {Human} {Factors} in {Computing} {Systems}},
	publisher = {Association for Computing Machinery},
	author = {Chiu, Yi-Yun and Hsieh, Wei-En and Lin, Wei-Chen and Tsai, Ning-Wei and Wu, Cheng-Han},
	month = apr,
	year = {2025},
	pages = {1--6},
}

@article{yaldiz_deepformabletag_2021,
	title = {{DeepFormableTag}: end-to-end generation and recognition of deformable fiducial markers},
	volume = {40},
	issn = {0730-0301},
	shorttitle = {{DeepFormableTag}},
	url = {https://dl.acm.org/doi/10.1145/3450626.3459762},
	doi = {10.1145/3450626.3459762},
	number = {4},
	urldate = {2025-08-07},
	journal = {ACM Trans. Graph.},
	author = {Yaldiz, Mustafa B. and Meuleman, Andreas and Jang, Hyeonjoong and Ha, Hyunho and Kim, Min H.},
	month = jul,
	year = {2021},
	pages = {67:1--67:14},
}

@inproceedings{silapasuphakornwong_embedding_2020,
	title = {Embedding {Information} in {3D} {Printed} {Objects} with {Curved} {Surfaces} {Using} {Near} {Infrared} {Fluorescent} {Dye}},
	language = {en},
	author = {Silapasuphakornwong, Piyarat and Torii, Hideyuki and Uehira, Kazutake and Chandenduang, Siravich},
	year = {2020},
    pages = {19--23},
    booktitle = {{MMEDIA} 2020, {The} {Twelfth} {International} {Conference} on {Advances} in {Multimedia}},
    publisher = {International Academy, Research, and Industry Association},
}

@inproceedings{gioberto_detecting_2013,
	address = {New York, NY, USA},
	series = {{ISWC} '13},
	title = {Detecting bends and fabric folds using stitched sensors},
	isbn = {978-1-4503-2127-3},
	url = {https://dl.acm.org/doi/10.1145/2493988.2494355},
	doi = {10.1145/2493988.2494355},
	urldate = {2025-08-05},
	booktitle = {Proceedings of the 2013 {International} {Symposium} on {Wearable} {Computers}},
	publisher = {Association for Computing Machinery},
	author = {Gioberto, Guido and Coughlin, James and Bibeau, Kaila and Dunne, Lucy E.},
	month = sep,
	year = {2013},
	pages = {53--56},
}

@inproceedings{luo_knitui_2021,
	address = {Yokohama Japan},
	title = {{KnitUI}: {Fabricating} {Interactive} and {Sensing} {Textiles} with {Machine} {Knitting}},
	isbn = {978-1-4503-8096-6},
	shorttitle = {{KnitUI}},
	url = {https://dl.acm.org/doi/10.1145/3411764.3445780},
	doi = {10.1145/3411764.3445780},
	language = {en},
	urldate = {2025-08-06},
	booktitle = {Proceedings of the 2021 {CHI} {Conference} on {Human} {Factors} in {Computing} {Systems}},
	publisher = {ACM},
	author = {Luo, Yiyue and Wu, Kui and Palacios, Tomás and Matusik, Wojciech},
	month = may,
	year = {2021},
	pages = {1--12},
}

@inproceedings{honnet_polysense_2020,
	address = {Honolulu HI USA},
	title = {{PolySense}: {Augmenting} {Textiles} with {Electrical} {Functionality} using {In}-{Situ} {Polymerization}},
	isbn = {978-1-4503-6708-0},
	shorttitle = {{PolySense}},
	url = {https://dl.acm.org/doi/10.1145/3313831.3376841},
	doi = {10.1145/3313831.3376841},
	language = {en},
	urldate = {2025-08-06},
	booktitle = {Proceedings of the 2020 {CHI} {Conference} on {Human} {Factors} in {Computing} {Systems}},
	publisher = {ACM},
	author = {Honnet, Cedric and Perner-Wilson, Hannah and Teyssier, Marc and Fruchard, Bruno and Steimle, Jürgen and Baptista, Ana C. and Strohmeier, Paul},
	month = apr,
	year = {2020},
	pages = {1--13},
}

@inproceedings{peng_intellitex_2024,
	address = {Honolulu HI USA},
	title = {{IntelliTex}: {Fabricating} {Low}-cost and {Washable} {Functional} {Textiles} using {A} {Double}-coating {Process}},
	isbn = {979-8-4007-0330-0},
	shorttitle = {{IntelliTex}},
	url = {https://dl.acm.org/doi/10.1145/3613904.3642759},
	doi = {10.1145/3613904.3642759},
	language = {en},
	urldate = {2025-08-06},
	booktitle = {Proceedings of the {CHI} {Conference} on {Human} {Factors} in {Computing} {Systems}},
	publisher = {ACM},
	author = {Peng, Yuecheng and Yan, Danchang and Chen, Haotian and Yang, Yue and Tao, Ye and Song, Weitao and Sun, Lingyun and Wang, Guanyun},
	month = may,
	year = {2024},
	pages = {1--18},
}

@inproceedings{dogan_standarone_2023,
	address = {New York, NY, USA},
	series = {{CHI} {EA} '23},
	title = {{StandARone}: {Infrared}-{Watermarked} {Documents} as {Portable} {Containers} of {AR} {Interaction} and {Personalization}},
	isbn = {978-1-4503-9422-2},
	shorttitle = {{StandARone}},
	url = {https://doi.org/10.1145/3544549.3585905},
	doi = {10.1145/3544549.3585905},
	urldate = {2025-08-05},
	booktitle = {Extended {Abstracts} of the 2023 {CHI} {Conference} on {Human} {Factors} in {Computing} {Systems}},
	publisher = {Association for Computing Machinery},
	author = {Dogan, Mustafa Doga and Siu, Alexa F. and Healey, Jennifer and Wigington, Curtis and Xiao, Chang and Sun, Tong},
	month = apr,
	year = {2023},
	pages = {1--7},
}

@inproceedings{devendorf_adacad_2023,
	address = {Hamburg Germany},
	title = {{AdaCAD}: {Parametric} {Design} as a {New} {Form} of {Notation} for {Complex} {Weaving}},
	isbn = {978-1-4503-9421-5},
	shorttitle = {{AdaCAD}},
	url = {https://dl.acm.org/doi/10.1145/3544548.3581571},
	doi = {10.1145/3544548.3581571},
	language = {en},
	urldate = {2025-08-05},
	booktitle = {Proceedings of the 2023 {CHI} {Conference} on {Human} {Factors} in {Computing} {Systems}},
	publisher = {ACM},
	author = {Devendorf, Laura and Walters, Kathryn and Fairbanks, Marianne and Sandry, Etta and Goodwill, Emma R},
	month = apr,
	year = {2023},
	pages = {1--18},
}

@article{weiser_computer_1999,
	title = {The computer for the 21st century},
	volume = {3},
	issn = {1559-1662},
	url = {https://dl.acm.org/doi/10.1145/329124.329126},
	doi = {10.1145/329124.329126},
	number = {3},
	urldate = {2025-08-04},
	journal = {SIGMOBILE Mob. Comput. Commun. Rev.},
	author = {Weiser, Mark},
	month = jul,
	year = {1999},
	pages = {3--11},
}

@inproceedings{liang_chic-marker_2024,
	address = {Aarhus Denmark},
	title = {Chic-{Marker}: {Fashionably} {Fusing} {Fiducial} {Markers} into {Apparel} and {Accessories}},
	isbn = {979-8-4007-0496-3},
	shorttitle = {Chic-{Marker}},
	url = {https://dl.acm.org/doi/10.1145/3639473.3665790},
	doi = {10.1145/3639473.3665790},
	language = {en},
	urldate = {2025-08-04},
	booktitle = {Proceedings of the 9th {ACM} {Symposium} on {Computational} {Fabrication}},
	publisher = {ACM},
	author = {Liang, Rong-Hao and Van Iterson, Hannah and Krueger, Holly and Toeters, Marina and Feijs, Loe},
	month = jul,
	year = {2024},
	pages = {1--15},
}

@inproceedings{preston_enabling_2017,
	address = {Edinburgh United Kingdom},
	title = {Enabling {Hand}-{Crafted} {Visual} {Markers} at {Scale}},
	isbn = {978-1-4503-4922-2},
	url = {https://dl.acm.org/doi/10.1145/3064663.3064746},
	doi = {10.1145/3064663.3064746},
	language = {en},
	urldate = {2025-08-04},
	booktitle = {Proceedings of the 2017 {Conference} on {Designing} {Interactive} {Systems}},
	publisher = {ACM},
	author = {Preston, William and Benford, Steve and Thorn, Emily-Clare and Koleva, Boriana and Rennick-Egglestone, Stefan and Mortier, Richard and Quinn, Anthony and Stell, John and Worboys, Michael},
	month = jun,
	year = {2017},
	pages = {1227--1237},
}

@inproceedings{devendorf_i_2016,
	address = {New York, NY, USA},
	series = {{CHI} '16},
	title = {"{I} don't {Want} to {Wear} a {Screen}": {Probing} {Perceptions} of and {Possibilities} for {Dynamic} {Displays} on {Clothing}},
	isbn = {978-1-4503-3362-7},
	shorttitle = {"{I} don't {Want} to {Wear} a {Screen}"},
	url = {https://dl.acm.org/doi/10.1145/2858036.2858192},
	doi = {10.1145/2858036.2858192},
	urldate = {2025-08-03},
	booktitle = {Proceedings of the 2016 {CHI} {Conference} on {Human} {Factors} in {Computing} {Systems}},
	publisher = {Association for Computing Machinery},
	author = {Devendorf, Laura and Lo, Joanne and Howell, Noura and Lee, Jung Lin and Gong, Nan-Wei and Karagozler, M. Emre and Fukuhara, Shiho and Poupyrev, Ivan and Paulos, Eric and Ryokai, Kimiko},
	month = may,
	year = {2016},
	pages = {6028--6039},
}

@inproceedings{hakkila_clothing_2017,
	address = {Stuttgart Germany},
	title = {Clothing integrated augmented reality markers},
	isbn = {978-1-4503-5378-6},
	url = {https://dl.acm.org/doi/10.1145/3152832.3152850},
	doi = {10.1145/3152832.3152850},
	language = {en},
	urldate = {2025-08-04},
	booktitle = {Proceedings of the 16th {International} {Conference} on {Mobile} and {Ubiquitous} {Multimedia}},
	publisher = {ACM},
	author = {Häkkilä, Jonna and Colley, Ashley and Roinesalo, Paula and Väyrynen, Jani},
	month = nov,
	year = {2017},
	pages = {113--121},
}

@inproceedings{chan_data_2017,
	address = {New York, NY, USA},
	series = {{UIST} '17},
	title = {Data {Storage} and {Interaction} using {Magnetized} {Fabric}},
	isbn = {978-1-4503-4981-9},
	url = {https://dl.acm.org/doi/10.1145/3126594.3126620},
	doi = {10.1145/3126594.3126620},
	urldate = {2025-07-30},
	booktitle = {Proceedings of the 30th {Annual} {ACM} {Symposium} on {User} {Interface} {Software} and {Technology}},
	publisher = {Association for Computing Machinery},
	author = {Chan, Justin and Gollakota, Shyamnath},
	month = oct,
	year = {2017},
	pages = {655--663},
}

@inproceedings{haberfellner_threads_2024,
	address = {New York, NY, USA},
	series = {{DIS} '24},
	title = {Threads of {Traceability}: {Textile} {IDs} in the {Fabric} of {Sustainable} {Fashion}},
	isbn = {979-8-4007-0583-0},
	shorttitle = {Threads of {Traceability}},
	url = {https://dl.acm.org/doi/10.1145/3643834.3661531},
	doi = {10.1145/3643834.3661531},
	urldate = {2025-07-30},
	booktitle = {Proceedings of the 2024 {ACM} {Designing} {Interactive} {Systems} {Conference}},
	publisher = {Association for Computing Machinery},
	author = {Haberfellner, Mira Alida and Zühlke, Samuel and Boess, Leopold and Probst, Kathrin and Haller, Michael},
	month = jul,
	year = {2024},
	pages = {3063--3078},
}

@inproceedings{pouta_woven_2022,
	address = {New York, NY, USA},
	series = {{DIS} '22},
	title = {Woven {eTextiles} in {HCI} — a {Literature} {Review}},
	isbn = {978-1-4503-9358-4},
	url = {https://dl.acm.org/doi/10.1145/3532106.3533566},
	doi = {10.1145/3532106.3533566},
	urldate = {2025-07-30},
	booktitle = {Proceedings of the 2022 {ACM} {Designing} {Interactive} {Systems} {Conference}},
	publisher = {Association for Computing Machinery},
	author = {Pouta, Emmi and Mikkonen, Jussi Ville},
	month = jun,
	year = {2022},
	pages = {1099--1118},
}

@inproceedings{zhu_research_2023,
	address = {Chengdu China},
	title = {Research on {Recognition} and {Registration} {Method} {Based} on {Deformed} {Fiducial} {Markers}},
	copyright = {https://www.acm.org/publications/policies/copyright\_policy\#Background},
	url = {https://dl.acm.org/doi/10.1145/3585542.3585543},
	doi = {10.1145/3585542.3585543},
	urldate = {2025-07-22},
	booktitle = {Proceedings of the 2023 7th {International} {Conference} on {Digital} {Signal} {Processing}},
	publisher = {ACM},
	author = {Zhu, Xiaoyu and Liu, Jiamin},
	month = feb,
	year = {2023},
	pages = {1--7},
}

@misc{yu_seampose_2024,
	title = {{SeamPose}: {Repurposing} {Seams} as {Capacitive} {Sensors} in a {Shirt} for {Upper}-{Body} {Pose} {Tracking}},
	shorttitle = {{SeamPose}},
	url = {http://arxiv.org/abs/2406.11645},
	doi = {10.48550/arXiv.2406.11645},
	urldate = {2025-07-22},
	publisher = {arXiv},
	author = {Yu, Tianhong Catherine and Manru and Zhang and He, Peter and Lee, Chi-Jung and Cheesman, Cassidy and Mahmud, Saif and Zhang, Ruidong and Guimbretière, François and Zhang, Cheng},
	month = jun,
	year = {2024},
	note = {arXiv:2406.11645 [cs]
version: 1},
}

@inproceedings{ray_capafoldable_2023,
	address = {New York, NY, USA},
	series = {{UbiComp}/{ISWC} '23 {Adjunct}},
	title = {Capafoldable: {Self}-tracking {Foldable} {Smart} {Textiles} {With} {Capacitive} {Sensing}},
	isbn = {979-8-4007-0200-6},
	shorttitle = {Capafoldable},
	url = {https://dl.acm.org/doi/10.1145/3594739.3610791},
	doi = {10.1145/3594739.3610791},
	urldate = {2025-07-21},
	booktitle = {Adjunct {Proceedings} of the 2023 {ACM} {International} {Joint} {Conference} on {Pervasive} and {Ubiquitous} {Computing} \& the 2023 {ACM} {International} {Symposium} on {Wearable} {Computing}},
	publisher = {Association for Computing Machinery},
	author = {Ray, Lala Shakti Swarup and Geißler, Daniel and Zhou, Bo and Lukowicz, Paul and Greinke, Berit},
	month = oct,
	year = {2023},
	pages = {197},
}

@inproceedings{li_e-textile_2024,
	address = {New York, NY, USA},
	series = {{TEI} '24},
	title = {E-textile {Sleeve} with {Graphene} {Strain} {Sensors} for {Arm} {Gesture} {Classification} of {Mid}-{Air} {Interactions}},
	isbn = {979-8-4007-0402-4},
	url = {https://dl.acm.org/doi/10.1145/3623509.3633374},
	doi = {10.1145/3623509.3633374},
	urldate = {2025-07-21},
	booktitle = {Proceedings of the {Eighteenth} {International} {Conference} on {Tangible}, {Embedded}, and {Embodied} {Interaction}},
	publisher = {Association for Computing Machinery},
	author = {Li, Yangfangzheng and Zhou, Yi and Shen, Cheng and Stewart, Rebecca},
	month = feb,
	year = {2024},
	pages = {1--10},
}

@inproceedings{dogan_infraredtags_2022,
	address = {New York, NY, USA},
	series = {{CHI} '22},
	title = {{InfraredTags}: {Embedding} {Invisible} {AR} {Markers} and {Barcodes} {Using} {Low}-{Cost}, {Infrared}-{Based} {3D} {Printing} and {Imaging} {Tools}},
	isbn = {978-1-4503-9157-3},
	shorttitle = {{InfraredTags}},
	url = {https://dl.acm.org/doi/10.1145/3491102.3501951},
	doi = {10.1145/3491102.3501951},
	urldate = {2025-07-04},
	booktitle = {Proceedings of the 2022 {CHI} {Conference} on {Human} {Factors} in {Computing} {Systems}},
	publisher = {Association for Computing Machinery},
	author = {Dogan, Mustafa Doga and Taka, Ahmad and Lu, Michael and Zhu, Yunyi and Kumar, Akshat and Gupta, Aakar and Mueller, Stefanie},
	month = apr,
	year = {2022},
	pages = {1--12},
}

@inproceedings{willis_hideout_2013,
	address = {New York, NY, USA},
	series = {{TEI} '13},
	title = {{HideOut}: mobile projector interaction with tangible objects and surfaces},
	isbn = {978-1-4503-1898-3},
	shorttitle = {{HideOut}},
	url = {https://doi.org/10.1145/2460625.2460682},
	doi = {10.1145/2460625.2460682},
	urldate = {2025-07-04},
	booktitle = {Proceedings of the 7th {International} {Conference} on {Tangible}, {Embedded} and {Embodied} {Interaction}},
	publisher = {Association for Computing Machinery},
	author = {Willis, Karl D. D. and Shiratori, Takaaki and Mahler, Moshe},
	month = feb,
	year = {2013},
	pages = {331--338},
}

@inproceedings{dogan_brightmarker_2023,
	address = {New York, NY, USA},
	series = {{UIST} '23},
	title = {{BrightMarker}: {3D} {Printed} {Fluorescent} {Markers} for {Object} {Tracking}},
	isbn = {979-8-4007-0132-0},
	shorttitle = {{BrightMarker}},
	url = {https://dl.acm.org/doi/10.1145/3586183.3606758},
	doi = {10.1145/3586183.3606758},
	urldate = {2025-07-04},
	booktitle = {Proceedings of the 36th {Annual} {ACM} {Symposium} on {User} {Interface} {Software} and {Technology}},
	publisher = {Association for Computing Machinery},
	author = {Dogan, Mustafa Doga and Garcia-Martin, Raul and Haertel, Patrick William and O'Keefe, Jamison John and Taka, Ahmad and Aurora, Akarsh and Sanchez-Reillo, Raul and Mueller, Stefanie},
	month = oct,
	year = {2023},
	pages = {1--13},
}

@inproceedings{rosner_spyn_2010,
	address = {New York, NY, USA},
	series = {{CHI} '10},
	title = {Spyn: augmenting the creative and communicative potential of craft},
	isbn = {978-1-60558-929-9},
	shorttitle = {Spyn},
	url = {https://doi.org/10.1145/1753326.1753691},
	doi = {10.1145/1753326.1753691},
	urldate = {2025-07-03},
	booktitle = {Proceedings of the {SIGCHI} {Conference} on {Human} {Factors} in {Computing} {Systems}},
	publisher = {Association for Computing Machinery},
	author = {Rosner, Daniela K. and Ryokai, Kimiko},
	month = apr,
	year = {2010},
	pages = {2407--2416},
}

@inproceedings{k_p_weaving_2023,
	address = {Cancun, Quintana Roo Mexico},
	title = {Weaving {Augmented} {Reality} {Markers}},
	isbn = {979-8-4007-0200-6},
	url = {https://dl.acm.org/doi/10.1145/3594739.3610786},
	doi = {10.1145/3594739.3610786},
	language = {en},
	urldate = {2025-07-02},
	booktitle = {Adjunct {Proceedings} of the 2023 {ACM} {International} {Joint} {Conference} on {Pervasive} and {Ubiquitous} {Computing} \& the 2023 {ACM} {International} {Symposium} on {Wearable} {Computing}},
	publisher = {ACM},
	author = {K. P., Nikita Menon and Lazaro Vasquez, Eldy S. and Curran, Hannah and De Koninck, Sasha and Devendorf, Laura},
	month = oct,
	year = {2023},
	pages = {310--314},
}

@inproceedings{feick_imprinto_2025,
	address = {Yokohama Japan},
	title = {Imprinto: {Enhancing} {Infrared} {Inkjet} {Watermarking} for {Human} and {Machine} {Perception}},
	isbn = {979-8-4007-1394-1},
	shorttitle = {Imprinto},
	url = {https://dl.acm.org/doi/10.1145/3706598.3713286},
	doi = {10.1145/3706598.3713286},
	language = {en},
	urldate = {2025-07-02},
	booktitle = {Proceedings of the 2025 {CHI} {Conference} on {Human} {Factors} in {Computing} {Systems}},
	publisher = {ACM},
	author = {Feick, Martin and Tang, Xuxin and Garcia-Martin, Raul and Luchianov, Alexandru and Huang, Roderick Wei Xiao and Xiao, Chang and Siu, Alexa and Dogan, Mustafa Doga},
	month = apr,
	year = {2025},
	pages = {1--18},
}

@article{narita_dynamic_2017,
	title = {Dynamic {Projection} {Mapping} onto {Deforming} {Non}-{Rigid} {Surface} {Using} {Deformable} {Dot} {Cluster} {Marker}},
	volume = {23},
	copyright = {https://ieeexplore.ieee.org/Xplorehelp/downloads/license-information/IEEE.html},
	issn = {1077-2626},
	url = {http://ieeexplore.ieee.org/document/7516689/},
	doi = {10.1109/TVCG.2016.2592910},
	language = {en},
	number = {3},
	urldate = {2025-08-11},
	journal = {IEEE Transactions on Visualization and Computer Graphics},
	author = {Narita, Gaku and Watanabe, Yoshihiro and Ishikawa, Masatoshi},
	month = mar,
	year = {2017},
	pages = {1235--1248},
}

@article{otsu_threshold_1979,
	title = {A {Threshold} {Selection} {Method} from {Gray}-{Level} {Histograms}},
	volume = {9},
	issn = {0018-9472, 2168-2909},
	url = {http://ieeexplore.ieee.org/document/4310076/},
	doi = {10.1109/TSMC.1979.4310076},
	language = {en},
	number = {1},
	urldate = {2025-09-10},
	journal = {IEEE Transactions on Systems, Man, and Cybernetics},
	author = {Otsu, Nobuyuki},
	month = jan,
	year = {1979},
	pages = {62--66},
}



\end{document}